\documentclass[
               twocolumn, 
               astrosymbooll,
               paper]{aastex701}
\usepackage[utf8]{inputenc}
\usepackage{amsmath}
\usepackage{xcolor}
\usepackage{appendix}
\usepackage{longtable}
\usepackage{natbib}
\usepackage{booktabs}
\graphicspath{{./}{figures/}}

\def\ie{{\it i.e.,}}
\def\eg{{\it e.g.,}}

\newcommand{\Ias}{SNe~Ia}

\newcommand{\Ibns}{SNe~Ibn}

\newcommand{\blIc}{SN~Ic-bl}

\newcommand{\abcsn}{\texttt{ABC-SN}}
\newcommand{\snid}{\texttt{SNID}}
\newcommand{\dash}{\texttt{DASH}}
\newcommand{\sniascore}{\texttt{SNIascore}}
\newcommand{\ccsnscore}{\texttt{CCSNscore}}

\newcommand{\gelato}{\texttt{GELATO}}
\newcommand{\ngsf}{\texttt{NGSF}}

\newcommand{\lsst}{\textit{LSST}}

\newcommand{\sedm}{\texttt{SEDM}}

\newcommand{\nSpec}{3574}
\newcommand{\nSpecfinal}{3395}

\newcommand{\nSpecCulled}{190}
\newcommand{\nSpecClipped}{179}

\newcommand{\numSNR}{\ensuremath{16}}
\newcommand{\numR}{\ensuremath{16}}
\newcommand{\numTotalModels}{$544$}  

\newcommand{\signal}{\texttt{signal}}
\newcommand{\noise}{\texttt{noise}}
\newcommand{\newnoise}{\texttt{new\_noise}}
\newcommand{\newspectrum}{\texttt{new\_spectrum}}

\newcommand{\Halpha}{{H$\alpha$}}

\newcommand{\SNR}{\ensuremath{\text{SNR}}}
\newcommand{\R}{\ensuremath{\ensuremath{R_\lambda}}}
\newcommand{\SNRnew} {\ensuremath{\text{SNR}_{\text{new}}}}
\newcommand{\Rnew}{$R_{\lambda\text{,new}}$}
\newcommand{\lname}[2]{#1 \textsc{#2}}

\newcommand{\spl}[1]{$\lambda #1$}

\newcommand{\vel}[1]{$#1 \textrm{ km s}^{-1}$}
\newcommand{\atomic}[2]{${}^{#2}$#1}

\newcommand{\msun}[1]{$#1 \textrm{ M}_\odot$}

\defcitealias{fortino_abc-sn_2026}{\scshape F26}

\shorttitle{How Low Can We Go?}
\shortauthors{Fortino et al.}

\begin{document}
\title{How Low Can We Go?\\Minimum Spectroscopic Requirements For Supernova Subtype Classification}
\correspondingauthor{Willow Fox Fortino}

\author[0000-0001-7559-7890]{Willow Fox Fortino}\email{fortino@udel.edu}
\affiliation{Department of Physics and Astronomy, University of Delaware, Newark, DE 19716, USA}

\author[0000-0003-1953-8727]{Federica B. Bianco}\email{fbianco@udel.edu}
\affiliation{Department of Physics and Astronomy, University of Delaware, Newark, DE 19716, USA}
\affiliation{Joseph R. Biden, Jr. School of Public Policy and Administration, University of Delaware, Newark, DE 19716, USA}
\affiliation{Data Science Institute, University of Delaware, Newark, DE 19716, USA}
\affiliation{Vera C. Rubin Observatory, Tucson, AZ, USA}

\author[0000-0001-7132-0333]{Maryam Modjaz}
\affiliation{Department of Astronomy, University of Virginia, 530 McCormick Road, Charlottesville, VA 22904, USA}
\email{mmodjaz@virginia.edu}

\author[0000-0001-6685-0479]{Thomas Matheson}
\affiliation{NSF NOIRLab, 950 N Cherry Ave, Tucson, AZ 85719, USA}
\email{tom.matheson@noirlab.edu}

\author[0000-0003-1671-8722]{Umer Zubair}
\affiliation{Department of Physics and Astronomy, University of Delaware, Newark, DE 19716, USA}
\affiliation{Department of Physics, West Chester University of Pennsylvania, West Chester, PA 19383, USA}
\affiliation{Department of Biomedical Engineering, West Chester University of Pennsylvania, West Chester, PA 19383, USA}
\email{umer@udel.edu}

\begin{abstract}
Millions of supernovae will be discovered with the Vera C. Rubin Observatory Legacy Survey of Space and Time (\lsst{}). As a result, spectrographs around the world will have to make difficult decisions about which supernova candidates receive spectroscopic follow-ups. This work identifies the minimum spectral resolution, $\R = \frac{\lambda}{\Delta  \lambda}$, as a function of signal-to-noise ratio (\SNR{}) at which spectral classification of supernova subtypes becomes impossible. We include supernova types Ia, Ia-91T, Ia-91bg, Iax, Ib, Ic, broad-lined Ic, IIb, IIP, and Ibn in this work. We produce a definition of \SNR{} based on specific lines for each SN subtype that allows us to generate homogeneous datasets at \numR{} different values of \R{} and \numSNR{} different \SNR's and we tested the classification performance of a recently developed deep-learning classifier, ABC-SN, on each \R{} and \SNR{} combination. We find that classification of supernova spectra into a refined taxonomy that separates, for example, between different subtypes of stripped envelope supernovae, is possible at low resolution and low SNR with no loss in model performance down to \R{} = 50 and \SNR = 5. Classification performance is only minimally impacted even as low as \R{} = 25. We hope that astronomers using the LSST alert stream, as well as designers of future instruments and observatories, will benefit from knowing what spectral resolution is necessary to classify a supernova for arbitrary \SNR{}.
\end{abstract}

\section{Introduction}\label{sec:intro}
\subsection{Supernovae}\label{sec:supernovae}
Stellar death can be violent. Some stars live quiet lives until fusion stops in their cores, while others end in a supernova (SN), an extraordinarily energetic and spectacular event. A typical type Ia SN releases $10^{51} \textrm{ ergs}$ of energy (\eg{} \citealt[][ch.~1]{athem_alsabti_handbook_2015}), and the extreme temperatures in excess of $10^9 \textrm{ K}$ are enough to fuse isotopes of titanium, vanadium, chromium, manganese, iron, cobalt, and nickel \citep[\eg{}][]{burbidge_synthesis_1957,wallerstein_synthesis_1997}. These elements, along with all of those made during the star's lifetime, are spread throughout the universe in these explosions, which are bright enough to outshine their own galaxy. In fact, SNe get their bright shine partially from the decay products of radioactive \atomic{Ni}{56} which can emit $10^{49} \textrm{ ergs}$ in only two weeks of the SN life \citep[\eg{}][]{colgate_early_1969, diehl_sn2014j_2015}.

\begin{table*}[t]
    \begin{tabular}{c|c|c|c|c|c}
    \textbf{Subtype} & \textbf{Thermonuclear} & \textbf{Core-Collapse} & \textbf{Stripped-Envelope} & \textbf{Interacting} & \textbf{Transition} \\ \hline
    Ia-norm  & x &   &   &   &   \\ \hline
    Ia-91T   & x &   &   &   &   \\ \hline
    Ia-91bg  & x &   &   &   &   \\ \hline
    Iax      & x &   &   &   &   \\ \hline
    Ib-norm  &   & x & x &   &   \\ \hline
    Ibn      &   & x & x & x &   \\ \hline
    IIb      &   & x & x &   & x \\ \hline
    Ic-norm  &   & x & x &   &   \\ \hline
    Ic-broad &   & x & x &   &   \\ \hline
    IIP      &   & x &   &   &   \\
    \end{tabular}
    \caption{The ten supernova subtypes used in this work, along with their general distinctions.}
    \label{tab:subtypes}
\end{table*}

Depending on the progenitor, the most common channels that lead to a supernova are a thermonuclear explosion or a core-collapse explosion. Thermonuclear supernovae occur when two carbon-oxygen (C/O) white dwarfs (WD) merge, or one accretes matter from a companion star until its mass is too great and overcomes the electron degeneracy pressure that keeps the WD from collapsing \citep{chandrasekhar_maximum_1931}. This picture of thermonuclear SNe is a vast simplification, and there is evidence for super- and sub-Chandrasekhar mass thermonuclear SNe. For a recent and thorough review on the topic, see \eg{} \citet{liu_type_2023}. Thermonuclear SNe have been identified as ``standardizable candles'' 
whereby the shape of the SN light curve can be related back to the SN intrinsic brightness \citep{1984SvA....28..658P, phillips1993absolute, 1997AAS...191.8504P}, making them invaluable tools for cosmology and leading to the discovery of an accelerating universe \citep{riess_observational_1998,perlmutter_measurements_1999,riess_precise_1996,tripp_using_1997}.

A core-collapse supernova (CCSN) occurs at the end of the life of massive stars \msun{>8} \citep[\eg{}][]{smartt_progenitors_2009} when they run out of fusion material. A balance between gravity and radiation pressure keeps the star stable, but when fusion stops, this balance breaks and the core of the star collapses \citep[\eg{}][]{couch_mechanisms_2017}. This happens in a fraction of a second and is followed by a rebounding explosion leading to the observable CCSN \citep[\eg{}][]{fryer_compact_2013}. Sometimes the progenitor of a CCSN may shed its outer envelope of hydrogen and helium, leading to SNe with weak or absent H and He lines; these are known as stripped-envelope SNe (SESNe) \citep[\eg{}][]{filippenko_optical_1997,modjaz_optical_2014}.

\citet{minkowski_spectra_1941} introduced the first SNe classification scheme by dividing the SNe that did not contain hydrogen (Type I) from those that did (Type II). Further `subtypes' have been defined since, including in \citet{blondin_determining_2007}, \citet{gal-yam_most_2019} \citet{silverman_berkeley_2012}, \citet{2019NatAs...3..717M}, and many more. One of the biggest challenges for a SN scientist is accurately classifying, or `typing,' SNe. This typically involves taking a spectrum of a SN and using the presence or absence of spectral lines to determine its subtype. In this work, we will exclusively focus on the optical range, which is where we can most easily collect SN spectra.

It must be noted that spectroscopic classification of SNe is complicated, compared to the spectral classification of stars, due to the blueshifting, broadening, and blending of spectral lines. The velocity of SN ejecta can range from \vel{3000 - 15,000} \citep{filippenko_optical_1997} and coupled with the fact that ejecta with both positive and negative line-of-sight velocities contribute to the radiative transfer, spectral features from SNe are not only redshifted by the cosmological expansion, but characteristically broad with varying degrees of P-cygni profiles, indicating moving photospheres.\footnote{For a thorough discussion of line broadening, see \eg{} an essay written by Stuart A. Sim in Ch.~31 of \textit{Handbook of Supernovae} \citep{athem_alsabti_handbook_2015} edited by Athem W. Alsabti and Paul Murdin.}. Note also that while all SNe have broadened (and blueshifted) lines compared to stars, we will also see that specific subtypes of SNe experience more broadening than the typical SN, since they are associated with more energetic events, and we refer to those as ``broad-line'' SNe, chief among them are broad-lined SN Ic, or SN Ic-bl or Ic broad \citep[\eg{}][]{iwamoto_peculiar_2000,woosley_supernova_2006,modjaz_early-time_2006}.

For this work, we make use of the ten SN subtypes shown in \autoref{tab:subtypes}. Beyond the three labels `thermonuclear,' `type II' and `stripped-envelope,' there are also `interacting' SNe where the ejecta from the SN explosion can interact with nearby static or slow-moving circumstellar material (CSM) of different compositions, causing distinctly narrow lines, \eg{} SNe IIn and Ibn, interacting with H-rich CSM and Icn interacting with C- and O-rich CSM. \citealt{chandra_multiwavelength_2025} offers a review on this topic. Furthermore, there are `transition' SNe with spectral features that change over time. All SN spectra evolve, especially between the photospheric and the nebular phase \citep[\eg{}][ch.~32]{athem_alsabti_handbook_2015}, but the transition subtype Type~IIb shows \textit{strong} hydrogen lines for the first few weeks, leading to a Type~II classification. Afterwards, the H lines fade, leading to a Type~Ib classification (\citealt{filippenko_supernova_1988}, but see also \citealt{liu_analyzing_2016}).\footnote{We record the age of a SN (also known as `phase' or `epoch') relative to the moment at which the explosion reaches its peak brightness.} Recently, even more kinds of transition SNe have been discovered, including with transitions before maximum light \citep[\eg{}][]{modjaz_shock_2009,chen_sn_2018}, and even SNe Ib may need time to develop their He lines \cite[\eg{}][]{williamson_optimal_2019}, but we will not discuss these here.

Many more types of SNe are not considered in this work, including Superluminous SNe \citep[SLSNe][]{gal2019most, kasen2011pair}, and a number of subtypes that emerge from surprising observations \citep[\eg{} 2018cow-like,][]{perley2019fast}, or from theoretical models eventually confirmed observationally \citep[\eg{} SN .Ia,][]{Shen_2010, PadillaGonzalez_2024}. 

\subsection{Spectroscopic Classification in the Era of Big Sky Surveys}\label{sec:rubin}
The Vera C. Rubin Observatory is a state-of-the-art optical observatory performing the Legacy Survey of Space and Time (\lsst{}), a ten-year photometric survey of the southern sky \citep{ivezic_lsst_2019}. The survey is estimated to result in $32$ \textit{trillion} observations of $17$ billion stars and $20$ billion galaxies. Among those observations will be $10$ million SNe \citep{lsst_science_collaboration_lsst_2009}. While this promises to revolutionize our understanding of stellar physics and SN Ia cosmology, it also poses an overabundance problem for the community. Similarly, the Argus Array \citep{law2022low} will observe the northern sky to a higher cadence, but lower limiting depth, leading to a similarly impressive number of discoveries.

The overwhelming quantity of incoming SNe motivates our work. As these surveys image the sky and discover hundreds of SNe candidates each night, spectroscopic follow-up classification is necessary to determine what science can be done with each SN. Therefore, spectrographs around the world will have to make difficult decisions about what to observe. Additionally, high signal-to-noise ratio (\SNR{}) spectra of distant objects may take an hour or more to observe. 

Today, the Zwicky Transient Facility (ZTF) \citep{bellm_zwicky_2018} is already delivering hundreds of SN observations per night. The ZTF Bright Transient Survey (BTS) uses custom SN classifiers.  One,
\sniascore{} \citep{fremling_sniascore_2021} is a machine-learning
 based classifier, designed exclusively for the \sedm{} and specifically designed
 for high SN~Ia classification purity.  It performs binary
 classification (Ia or non-Ia) and regresses SN redshift.  \ccsnscore{}
 \citep{sharma_ccsnscore_2025} is designed to classify subtypes of Ib,
 Ic, and II SNe on \sedm{} data.  Most recently, the ZTF team produced a
 multimodal neural network classifier that uses spectra, images, and
light curves \citep{junell2026applying}, that includes a convolutional
neural network, \texttt{SpectraNet}, for \sedm{} spectra classification
\citep{xu2025applecider}.

Most recently, the ZTF team produced a multimodal neural network classifier that uses spectra, images, and light curves \citep{junell2026applying}, which includes a convolutional neural network, \texttt{SpectraNet}, for \sedm{} spectra classification \citep{xu2025applecider}. 

The success of the ZTF SN surveys demonstrates vividly that SN classification at low resolution is indeed possible and optimal in the high-discovery volume present in astrophysical research. Higher-resolution spectrographs can then be reserved for detailed studies of specific SNe of particular interest.

However, it is currently unclear what combination of \SNR{} and spectroscopic resolution, \R{}, is needed to classify SNe accurately. There is no systematic evaluation of SN classification performance as a function of \SNR{} and \R{} that could inform future spectrograph design and observation strategy. This is the goal of our work.

\subsection{Modern Classification Techniques} \label{sec:classification}
A spectrum is necessary to classify most SN subtypes. The most used method for determining SNe type is SuperNova IDentification (\snid{}) \citep{blondin_determining_2007}, which implements the cross-correlation technique proposed by \citet{tonry_survey_1979}. \snid{} relies on a database of high \SNR{} template spectra that have already been classified by either an expert human, \snid{}, or some other automated means. The program calculates the correlation between the unknown input spectrum and all template spectra. The input spectrum's subtype is then taken to be the subtype of the template spectrum that has the highest correlation with the input while also providing redshift and phase estimates based on the matching template. All spectra used in this analysis were originally classified using \snid{}.

Template-matching methods like \snid{} (but also its successor \texttt{SNID-SAGE} \citealt{Stoppa2026}, Superfit \citealt{howell_gemini_2005}, Next Generation SuperFit --- \ngsf{} ---  \citealt{kim_how_2024}, GEneric cLAssification TOol --- \gelato{} --- \citealt{harutyunyan_esc_2008}) benefit from being highly interpretable. That is, a classification from \snid{} can be explained by examining which specific template spectrum was most correlated with the input. However, their computational cost is on the order of seconds per spectrum with a template library of 1,000 spectra \citep[\eg{} for SNID $6s \times (\mathrm{CPU} / 2.86~GHz) \times (N_\mathrm{templates} / 1000)$,][]{blondin_determining_2007}, and scales linearly with the number of templates - expensive when compared to neural network inference times (\eg{}  $128 \mu s$  per spectrum for \citetalias{fortino_abc-sn_2026}). Furthermore, they are susceptible to spurious correlation (\eg{} between galaxy or telluric features contaminating the spectra, or instrumental signatures) and typically require human validation.

For example, a key hyperparameter in SNID is \texttt{rlap}, defined as $\texttt{rlap} = r \times \texttt{lap} < \texttt{rlapmin}$, where $r$ is the correlation coefficient and \texttt{lap} is the natural logarithm of the ratio of the overlap wavelength range between the spectra. The threshold \texttt{rlapmin} allows the user to reject spectra with unreliable matches, but when used with rest-frame spectra it reduces one’s ability to classify high-redshift SNe and requires a deep understanding of the model to set appropriate values.\footnote{The \snid{} manual states ``Make sure you know what you’re doing before changing the value of \texttt{rlapmin}! Correlations with $\texttt{rlap} < \texttt{rlapmin}$ are not used for type, redshift, and age statistics.''}

Machine learning presents an alternative to the template-matching, cross-correlation classification paradigm. The first deep-learning SN Classifier appeared in 2019 \citep[\dash,][]{muthukrishna_dash_2019}. The Attention-Based Classifier for Supernovae \citep[\abcsn{},][hereafter \citetalias{fortino_abc-sn_2026}]{fortino_abc-sn_2026} is a deep learning classification model that employs the attention mechanism as demonstrated with the Transformer \citep{vaswani_attention_2017}. The Transformer has been shown to learn language (in a machine learning sense) very well, as is evidenced by the generally convincing, human-like text that large language models like GPT \citep{devlin2019bert}, Claude \citep{anthropic2025claude}, and DeepSeek \citep{bi2024deepseek} can create. For the same reason that transformers can recognize and relate the complex web of language and syntax to understand and construct new sentences, \abcsn{} uses the transformer to understand the relationship among lines in a SN spectrum.

Our goal with this work is to use \abcsn{}, along with a highly curated dataset of SN spectra originally comprising the \snid{} template library (and thus accurately classified for the purpose), to measure the impact on classification power as the \R{} and \SNR{} of the dataset change. Section~\ref{sec:data} discusses the origin and composition of the dataset as well as the data preprocessing steps we took to make the data ready for machine learning. Section~\ref{sec:method} explains the methods we used to change the \R{} and \SNR{} of the data, as well as the details of how we re-trained \abcsn{}. Section~\ref{sec:results} explains the resulting performance of \abcsn{}, and in Section~\ref{sec:conclusion} we conclude.

\section{Data}\label{sec:data}
To test the response of an automated SN spectra classifier to data\SNR{} and \R{} we use the recently developed ABC-SN. This model has demonstrated state-of-the-art performance at $\R=100$, and we have complete control and understanding of its structure and performance since it was developed by our group.

In this work (as well as in the original \citetalias{fortino_abc-sn_2026}), we use a physically motivated, restricted taxonomy that includes subtypes of normal luminosity SNe (\ie{} not superluminous like SLSNe or subluminous like SNe .Ia) and for which there have been enough observed examples to construct a machine learning dataset. Our taxonomy is more detailed than the current capabilities of photometric classifiers \citep{boone_avocado_2019, karpenka2013simple, qu_scone_2021, moller_supernnova_2020}, which typically combine stripped envelope SNe into a SN Ib/c class, different subtypes of unusual SNe Ia into Ia-pec, \emph{etc}. While there is increasing interest in classifying different transients like Active Galactic Nuclei (AGN) or Tidal Disruption Events (TDE), we limit this work to SNe to establish a baseline and maintain consistency with previous work and our reference models (in particular, \dash{} and \abcsn{}). We leave the extension of our results to different transient classes for future work. 

The data used in this analysis are the same data that were used to train \abcsn{}, but with some updates and modifications described below. This dataset offers specific desirable properties for this work. The spectra are originally sourced from an augmented \snid{} template library; \snid{} \citep{blondin_using_2006} is the most common SN spectrum classifier aside from instrument- or facility-specific models. Since the spectra in our dataset were all designed to be used as \snid{} templates, they are uniform in their preprocessing and generally high \SNR{}. We note here that the definition of \SNR{} for SN spectra is not uniform in the literature. We define a specific way to measure the \SNR{} of a SN spectrum in \autoref{sec:snr}, and all references to \SNR{} in our work refer to that measurement. By our method, the median \SNR{} of the spectra in our dataset is $\approx20$, with an inter-quartile range of $\mathrm{IQR}\approx10
-38.79$.

\subsection{Data Provenance}\label{sec:provenance}
A detailed description of the original data and all data processing steps is offered in \citetalias{fortino_abc-sn_2026}, but we summarize it here for the reader's convenience.

All spectra are originally extracted from the \snid{} template library and were preprocessed according to \citealt{blondin_determining_2007}. Spectra are retrieved from many original sources including: \snid{} templates from \citet{blondin_determining_2007, blondin_spectroscopic_2012}, a catalog of stripped envelope supernovae (SESNe) \citep{liu_supernova_2015, modjaz_optical_2014, modjaz_spectral_2016, liu_analyzing_2016}, a catalog of spectra from the Berkeley SN Ia Program (BSNIP) \citep{silverman_berkeley_2012}, spectra from the \texttt{SUSPECT} public archive \citep{yaron_wiserepinteractive_2012}, and spectra from the CfA Supernova Archive and the CfA Supernova Program \citep{matheson_optical_2008, blondin_spectroscopic_2012}.

\begin{table*}
    \centering
    \begin{tabular}{c|cccccc}
         & \citetalias{fortino_abc-sn_2026} & Manual  & Outliers & Training Set & Test Set & Augmented \\
         
         & & Review & Removed & (\#SNe) & (\#SNe) &  Training Set\\
         \hline
         Ia-norm  & 2114 & 2034 & 1988 & 930 (119) & 1058 (119) & 1860 \\
         Ia-91T   & 348 & 339 & 321 & 186 (14) & 135 (13) & 1860\\
         Ia-91bg  & 232 & 222 & 201 & 90 (14) & 111 (14) & 1890 \\
         Iax      & 62 & 60 & 60 & 29 (2) & 31 (1) & 1885 \\
         \hline
         Ib-norm  & 211 & 198 & 189 & 111 (11) & 78 (11) & 1887 \\
         Ibn      & 27 & 26 & 11 & 8 (2) & 3 (1) & 1864 \\
         IIb      & 232 & 205 & 181 & 70 (10) & 111 (9) & 1890\\
         \hline
         Ic-norm  & 206 & 180 & 162 & 117 (9) & 45 (9) & 1872 \\
         Ic-broad & 228 & 210 & 205 & 138 (10) & 67 (10) & 1932 \\
         \hline
         IIP      & 104 & 100 & 77 & 39 (3) & 38 (3) & 1872 \\
    \end{tabular}
    \caption{The number of supernova spectra (and the number of distinct supernovae in parentheses) at each data processing stage. See \autoref{sec:preprocessing} for details and a description for the applied augmentation techniques.}
    \label{tab:spectra}
\end{table*}

\subsection{Data Composition}\label{sec:composition}
Each spectrum starts at a spectral resolution of $\R{}=738$ and is defined on the wavelength range $2,500-10,000 \text{ \AA}$.\footnote{Most spectra do not have flux observations throughout the entire wavelength range so this range will be cropped in later steps.} The dataset begins with spectra from 17 SN subtypes, though spectra corresponding to seven of those subtypes were removed in the training of \abcsn{}: Ia-csm, Ia-pec, Ib-pec, Ic-pec, IIL, IIn, and II-pec because there were insufficient spectra or insufficient supernovae to train a machine-learning classifier (see section~3 and table~1 in \citetalias{fortino_abc-sn_2026}). 

In particular, type IIn are a relatively common SN subtype of SNe, contributing to $\sim7\%$ of hydrogen-rich SNe, $\sim5\%$ of all core collapse SNe \citep{shivvers2017revisiting}, but they are not considered here because not enough supernovae and spectra are available in our sample to generate training and testing examples. We recognize this as a weakness of our result, particularly because the IIn SNe are characterized by narrow lines that may be lost at low resolution. However, we note that we do classify SN Ibn with ABC-SN with high accuracy at $\R{}=100$ (precision $Pr=100\%$ and recall $Re=85.7\%$, and degrading with lower \R{} at the same rate as for other subtypes). We further note that SN 91T and 91bg, both included in our taxonomy, are respectively over- and underluminous SN Ia subtypes with classification partially based on photometry. Here, we rely on the expertise of the scientists who compiled the \snid{} sample for their classification, but we note that \abcsn{} achieves relatively high precision and recall on both subtypes from the spectra alone: $Pr=83.2\%$ and $Pr=82.4\%$ and $Re=64.7\%$ and $Re=76.1\%$, respectively.

We also restrict the spectral phase to between $-20$ and $50$ days. Early and late classifications are particularly complex: early spectra have rapidly evolving spectral characteristics, and late spectra show features specific to the nebular phase and are systematically fainter. These initial cuts leave us with $3764$ spectra from $498$ unique SNe.

As argued in \citetalias{fortino_abc-sn_2026}, while not complete, this taxonomy is physically motivated, it is refined,  including a subdivision into subtypes where those subtypes are well-defined phenomenologically (as opposed to generic labels assigned to objects that do not fit other classes, such as `peculiar'), and it avoids the inclusion of subtypes that would increase noise and reduce the reliability of our results.

\subsection{Data Preprocessing} \label{sec:preprocessing}
\snid{} has its own series of preprocessing steps that are designed to enforce homogeneity for its analyses. Those preprocessing steps are described in detail in \citet{blondin_determining_2007}. The result is a de-redshifted, continuum-removed spectrum, re-binned to the spectral resolution of $\R=738$. Note that this spectral resolution is considered high for SN classification purposes, where the spectral features are broad because they are generated by high-velocity ejecta (see \autoref{sec:intro}).

To achieve our goal of measuring classification performance as a function of \SNR{} and \R{}, we simulate several datasets of spectra at various combinations of \SNR{} and \R{}. The following is a list of the preprocessing steps that our data experiences, starting from its retrieval and compilation and ending with its use in re-training \abcsn{}.

Each spectrum is first normalized to mean 0 and standard deviation 1. The initial \SNR{} of each spectrum is measured; the \SNR{} measurement algorithm is described in \autoref{sec:snr}. During this step, \nSpecCulled{} spectra are removed as they were found to have missing data between $4,500\text{ \AA}$ and $7,000\text{ \AA}$ (see Appendix). We can now begin to artificially change the \SNR{} of each spectrum in the dataset to \SNRnew{}.

The signal and noise arrays (\signal{} and \noise{}, respectively) are separated for each spectrum (as described in \autoref{sec:snr}). We can now construct a new spectrum, $\newspectrum = \signal + \newnoise$, such that the \SNR{} of \newspectrum{} is \SNRnew{}. We remeasure the \SNR{} of \newspectrum{} to ensure that it is close to the desired value of \SNRnew{}; \nSpecClipped{} spectra were removed in this step. Specifically, we removed spectra that fell into the 95th percentile of the re-measured \SNR{} distribution for $\SNRnew=100$. Thus, we refer to these spectra as ``outliers'' (see \autoref{tab:spectra} and Appendix \autoref{tab:clipped_spectra}). This leaves us with \nSpecfinal{} spectra in our dataset.

The spectra are split between a training and a test set (\autoref{tab:spectra}). Note that in the case of a spectral library where multiple spectra come from the same SN, this step requires significant curation, as described in \citetalias{fortino_abc-sn_2026}, to ensure all the spectra for each SN are in either the training or the test set in order to avoid data leakage. Data augmentation techniques are applied to the training set to achieve class balance (\autoref{sec:retrain}). The wavelength range is clipped to $4,500-7,000\text{ \AA}$ (\autoref{sec:retrain}), and the resolution of the spectra is lowered from $\R{} = 738$ to \Rnew{}. With these \numTotalModels{} datasets in hand, we can retrain \abcsn{} on the \SNRnew{}, \Rnew{} dataset. 

This concludes the data preprocessing steps we took to build each `dataset' of \nSpecfinal{} spectra, each with their \R{} lowered to \Rnew{} (\numR{} \R{} values) and their \SNR{} changed to \SNRnew{} (\numSNR{} \SNR{} values). Additionally, we  on \numR{} datasets at the original \SNR{} as a validation, where we expect to recover the performance measured in \citet{fortino_abc-sn_2026}. We performed a two-fold cross validation split, which doubles the number of datasets, leading to a total of \numTotalModels{} datasets. Finally, the data is ready to be used in a machine learning classifier.

\section{Methodology}\label{sec:method}
Our goal is to measure the performance of an automated SN classifier as a function of \SNR{} and \R{}. Below we describe the processing steps to produce derived datasets at varying \SNRnew{} and \Rnew{}, and the retraining of our chosen baseline classifier, \abcsn{}.

\subsection{Changing resolution}\label{sec:r}
\begin{figure}
    \centering
    \includegraphics[width=1\linewidth]{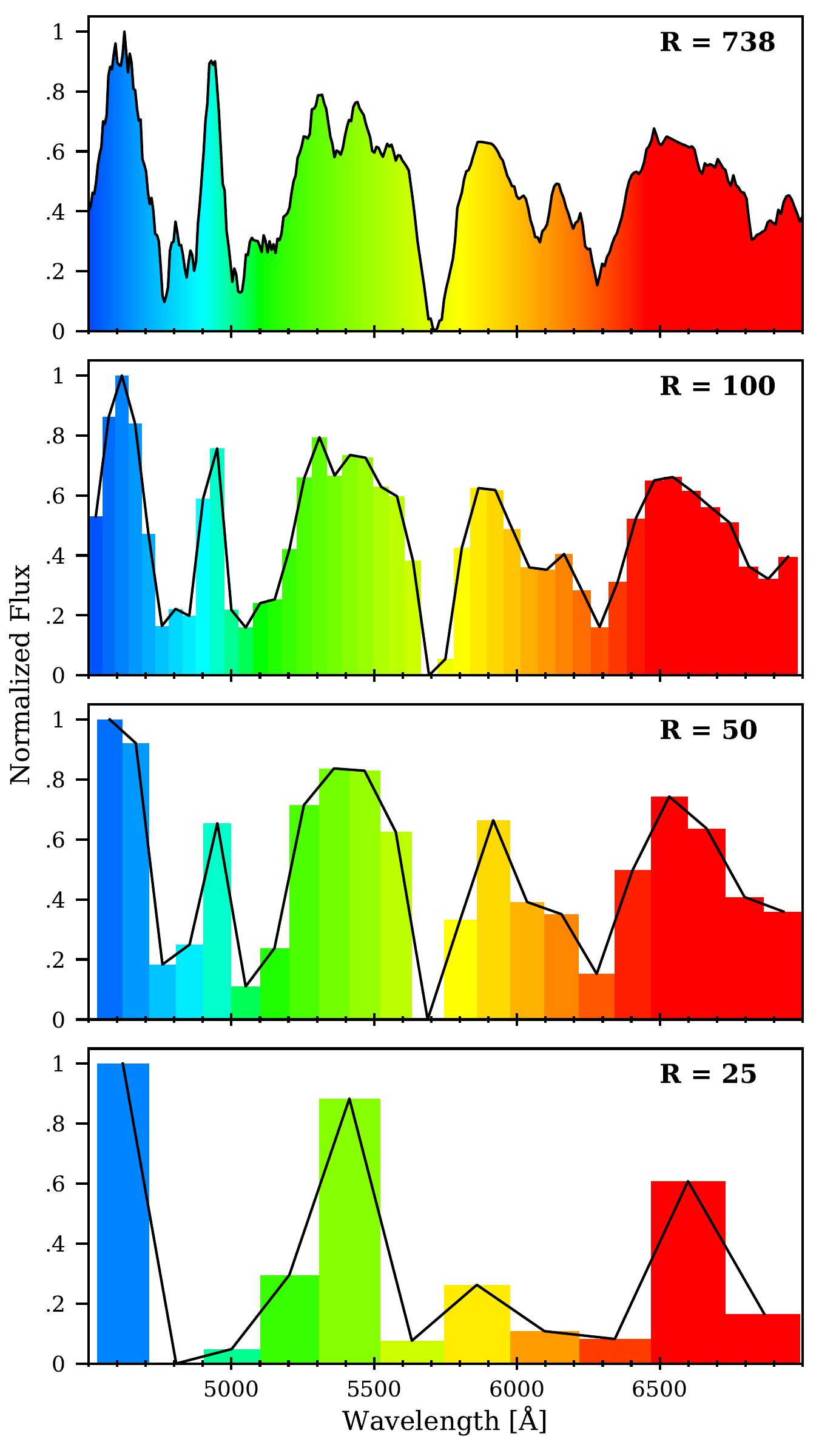}
    \caption{Type Ic Supernova 1990b at 4 days after peak brightness shown in four different spectral resolutions. $\R=738$, the resolution of the data as we retrieved it from \snid{} templates. $\R=100$, the resolution of the \sedm{} and the resolution that the original \abcsn{} was trained at. $\R=50$, the lowest resolution at which we measure no \abcsn{} performance loss for any \SNR. $\R=25$, below this resolution classification declines rapidly. Here the spectra were normalized to be between 0 and 1 for visualization purposes while \abcsn{} trains on standardized data.}
    \label{fig:example_specs}
\end{figure}

We describe the process of reducing spectral resolution \R{} in detail in \citetalias{fortino_abc-sn_2026}. In brief, we convolve the original spectrum with a Gaussian with a standard deviation $\sigma$ that is a function of bin-width. That is, $G(\mu, \sigma) = G(\lambda_i, \sigma(\Delta \lambda_i))$ where $G$ is a Gaussian function with mean $\mu$ and standard deviation $\sigma$, $\lambda_i$ is the wavelength of the center of bin $i$, and $\Delta \lambda_i$ is the size of bin $i$. We choose $\textrm{FWHM} = \Delta \lambda_i R_{\text{original}} / R_{\text{lower}}$. We can then linearly interpolate the convolution (which is defined on the original spectrum's wavelength bins) to the lower resolution wavelength array which simulates what a low-resolution spectrograph might observe. \autoref{fig:example_specs} shows an example of a spectrum at the original $\R = 738$, the resolution of the \sedm{} $\R = 100$, $\R=50$, and $R=25$ (which will be identified as critical values in \autoref{sec:conclusion}). For training and prediction, we do not linearly interpolate the convolution in order to retain the most information content in each spectrum.

\subsection{Changing \SNR }\label{sec:snr}
In order to investigate the combined effects of spectral resolution and \SNR{} on classification performance, we developed a novel, semi-automated algorithm to first measure, then artificially change, the \SNR{} of every spectrum in our dataset. While the process of lowering spectral resolution is well defined (see \autoref{sec:r} and \citetalias{fortino_abc-sn_2026}), there is no analogous standard method for modifying the \SNR{} of a SN spectrum. This begins with a robust measurement of the \SNR{}, which is itself neither trivial nor standardized.

The basic definition of \SNR{} is, trivially, $\SNR = S/N$, where $S$ is the amplitude of the signal and $N$ is the standard deviation of the noise. The noise comes from a variety of sources, including shot noise, detector read noise, sky noise, galaxy contamination, etc. For a SN spectrum, the signal is the flux in a spectral line. Despite the simple definition, actually measuring $S$ and $N$ represents significant challenges for SN spectra, and we find that there is neither a globally accepted definition of SN spectrum SNR, nor a standardized way to measure it for a given spectrum. 



Measuring the signal of a spectrum is significantly more involved because in a SN spectrum, each line has a different strength, and quantifying the `strength' of irregularly shaped, broad spectral features is not easy. One method, inherited from stellar spectroscopy, involves calculating the pseudo-equivalent width (pEW):

\begin{equation}
    \text{pEW} = \int_{\lambda_a}^{\lambda_b} 1-\frac{f_s(\lambda')}{f_{pc}(\lambda')} d\lambda'
\end{equation}

\noindent where $(\lambda_a,\lambda_b)$ is the range of the spectral feature, $f_s(\lambda)$ is the signal flux of the feature and $f_{pc}(\lambda)$ is the pseudo-continuum \citep{hoffmann2014non}. As its name suggests, the pseudo-continuum is a stand-in for a background, line-free continuum region. The pseudo-continuum is defined as a straight line between the bounds of the spectral feature. Thus, the pEW equation is the integral over the fractional change of the spectral feature relative to the pseudo-continuum.

In this work, we define $S$ based on the area of the spectral feature relative to the pseudo-continuum, as in \citet{folatelli_spectral_2004}, but we also divide by the total width of the feature.

\begin{equation}
    S = \frac{1}{\lambda_b - \lambda_a}\int_{\lambda_a}^{\lambda_b} |f_{pc}(\lambda') - f_s(\lambda')| d\lambda'.
    \label{eq:s}
\end{equation}

\begin{figure*}[!hp]
    \centering
    \includegraphics[width=1\linewidth]{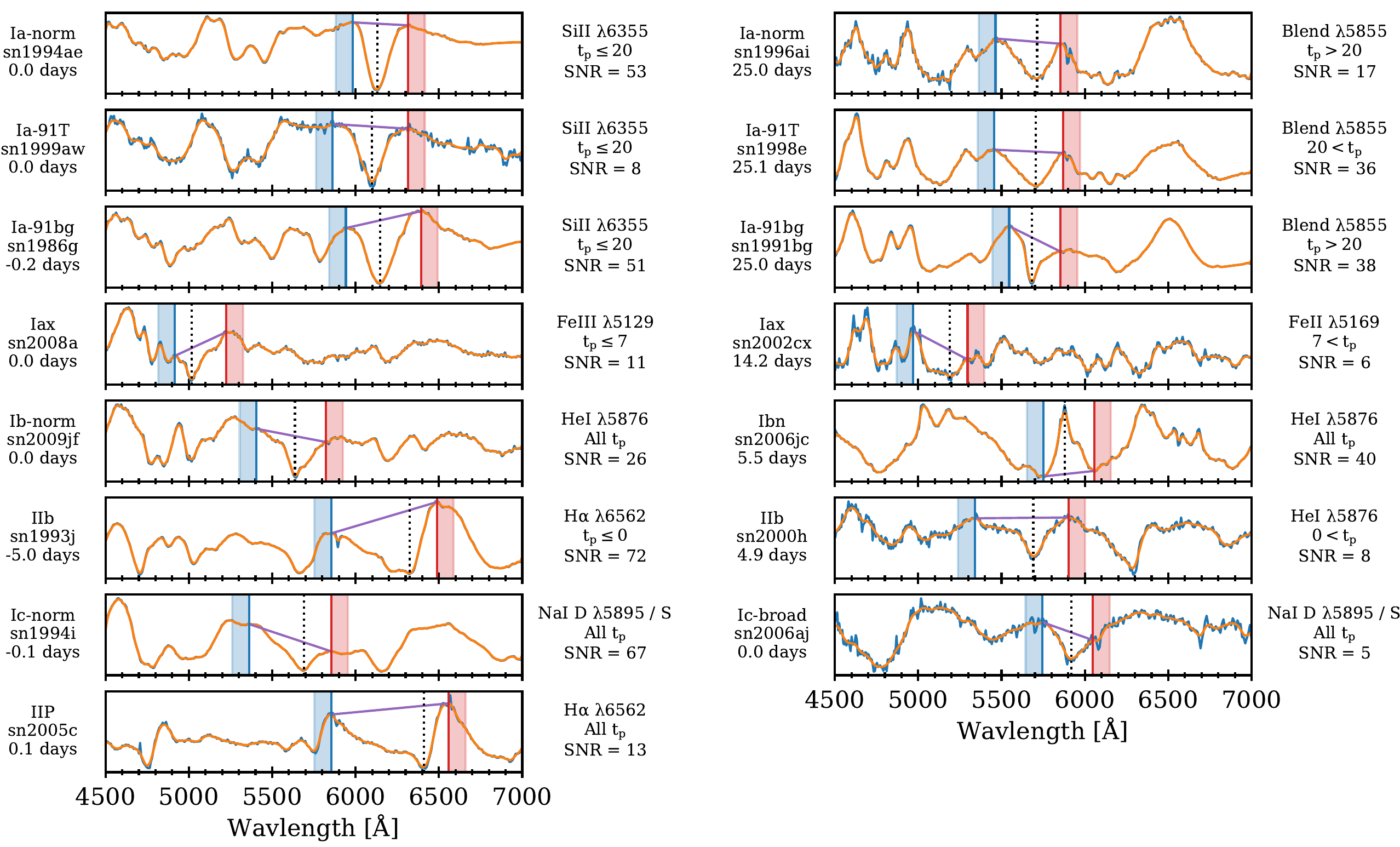}
    \caption{An example spectrum is shown for each feature that we measure for each subtype. The extracted \signal{} is plotted in orange on top of the \texttt{spectrum} in blue. The black dotted vertical line shows the extremum corresponding to the observed feature. The blue and red vertical lines correspond to the shoulders of the feature, and the blue and red shaded regions correspond to the wavelengths where the noise, $N$, is evaluated. The purple line represents the pseudo-continuum. All spectra are plotted at $\R=738$ and the measured \SNR{} of each spectrum is annotated.}
    \label{fig:features}
\end{figure*}

By dividing by the width of the feature, the resulting quantity will have units of normalized flux. Dividing by the width of the line effectively penalizes the \SNR\ of broad features, which are typical of spectral types like \blIc, but this reflects the more difficult challenge models face in classifying spectra with features that are more easily confused with the complex underlying continuum. \autoref{fig:features} shows several examples of SN spectra and the steps described above to measure the signal $S$ and the noise $N$ (described in \autoref{sec:snr}), using subtype-specific features as described in \autoref{sec:features}.

\subsubsection{Spectral Features}\label{sec:features}
In order to measure the \SNR{} of a SN spectrum, we must first identify which spectral feature we will measure. As discussed in \autoref{sec:supernovae}, SNe are classified based on the observed spectral features (\ie{} the chemical composition) and the phase of the SN. Ideally, there would exist one strong feature present in all subtypes at all phases that we could use to measure \SNR{}, but we are not so lucky. Additionally, we restrict each spectrum to the phase range $\lambda\in[4,500, 7,000]\text{ \AA}$, which means that many of the features common in SNe (\eg{} oxygen \spl{7774} or the Ca NIR triplet) are not available to us. Nonetheless, we identified eight distinct features that cover all ten subtypes and our chosen phase range of $t\in[-20, 50]\text{ days}$ as shown in \autoref{tab:features} and \autoref{fig:features}.

\begin{table}[t]
    \centering
    \begin{tabular}{c|c|c}
        \textbf{Subtype} & \textbf{Phase Range} & \textbf{Line} \\ \hline
        Ia-norm  & $[-20, 20]$ & \lname{Si}{ii} \spl{6355}      \\ 
        Ia-norm  & $(20, 50]$  & Blend \spl{5855}               \\ \hline
        Ia-91T   & $[-20, 20]$ & \lname{Si}{ii} \spl{6355}      \\
        Ia-91T   & $(20, 50]$  & Blend \spl{5855}               \\ \hline
        Ia-91bg  & $[-20, 20]$ & \lname{Si}{ii} \spl{6355}      \\
        Ia-91bg  & $(20, 50]$  & Blend \spl{5855}               \\ \hline
        Iax      & $[-20, 7]$  & \lname{Fe}{iii} \spl{5129}     \\
        Iax      & $(7, 50)$   & \lname{Fe}{ii} \spl{5169}      \\ \hline
        Ib-norm  & $[-20, 50]$ & \lname{He}{i} \spl{5876}       \\ \hline
        Ibn      & $[-20, 50]$ & \lname{He}{i} \spl{5876}       \\ \hline
        IIb      & $[-20, 0]$  & \Halpha{} \spl{6563}           \\
        IIb      & $(0, 50]$   & \lname{He}{i} \spl{5876}       \\ \hline
        Ic-norm  & $[-20, 50]$ & \lname{Na}{i} D \spl{5895} / S \\ \hline
        Ic-broad & $[-20, 50]$ & \lname{Na}{i} D \spl{5895} / S \\ \hline
        IIP      & $[-20, 50]$ & \Halpha{} \spl{6563}
    \end{tabular}
    \caption{For each subtype, we identify a prominent spectral feature that we can base our \SNR{} measurement on. Some subtypes require that we find two features depending on the phase. The \lname{He}{i} line is measured in emission for \Ibns{}, but all other features are measured in absorption. }
    \label{tab:features}
\end{table}

Supernovae Type Ia are well known for having a strong \lname{Si}{ii} feature at \spl{6355} which fades after a few weeks, so we use that for early time Ia-norm, Ia-91T and Ia-91bg spectra. In our wavelength range, there are not many consistently strong features after \lname{Si}{ii} fades, but we identified a blended feature by eye, around $500\text{ \AA}$ bluer than \lname{Si}{ii}, that becomes stronger as \lname{Si}{ii} gets weaker \citep{filippenko_optical_1997}. These features in SN Iax are comparatively weaker, so we decided not to search for the \lname{Si}{ii} feature for this subtype (a full discussion of \lname{Si}{ii} line velocity in SN Iax can be found in \citealt{foley_type_2013}). \citet{li_sn_2003} identifies two prominent iron features for us that we look for in SNe~Iax.

Lacking both \lname{Si}{ii} and \lname{He}{i}, for Ic (or Ic-norm) and Ic-bl we look to the \lname{Na}{i} D feature, which is reasonably prominent throughout our phase range \citep{filippenko_optical_1997}. We note, however, that this feature may not be pure \lname{Na}{i} D. Rather, it has been shown to be blended in SN 1994I with \lname{He}{i} \spl{5876} (\eg{} \citealt{dessart2012nature, sauer2006properties, patat2001metamorphosis, 1996ApJ...462..462C}) and \citet{williamson_modeling_2021} note that this line also includes sulfur  \lname{S}{ii} \spl{5640} and \lname{S}{ii} \spl{5606} (we will thus label it \lname{Na}{i} D \spl{5895} / S in \autoref{tab:features} and \autoref{fig:features}).

Lastly, SNe~IIP show strong \Halpha{} lines throughout their evolution, making it a straightforward choice \citep{filippenko_optical_1997}. Overall, we attempted to pick features that were shown in the literature to be emblematic of that subtype, rather than the strongest feature at any given phase.

\subsubsection{Separating Signal and Noise}
The first step in our process is to obtain a measurement of the ``signal'' and ``noise'' arrays (\signal{} and \noise{}) at their original resolution ($\R = 738$).\footnote{Note: this resolution is itself modified from the original raw data in the \snid{} preprocessing steps; see \autoref{sec:data}.}

Noise in astrophysical spectra is complex and, importantly, varies slowly with wavelength. The most convenient and effective method for calculating the noise of a spectrum would be to start with the raw 2D spectrum and propagate uncertainties at each stage of data reduction. Ideally, uncertainty arrays would be produced during data reduction and made available along with the processed spectrum and typically include removal of Poisson and CCD readout noise, and other sources of noise that have subdominant impacts \citep{horne1986optimal}. If the noise arrays are not made available, deriving them this way would require access to raw spectra, which is rarely available to astronomers if they were not the PI of the observing programs that collected the spectra, and were not available to us for our target dataset. Without the raw spectrum, the noise and signal may be separated by smoothing over the noise in the spectrum.

To produce the \noise{} then, we first standardize each spectrum to a $[0, 1]$ range and smooth it with a Gaussian filter. While specific methods have been described in the literature to separate noise and signal in SN spectra \citep[\eg{}][]{liu_analyzing_2016, blondin2006using, blondin_determining_2007}, we find that a simple Gaussian filter leads to more stable results for these spectra that are already preprocessed and continuum removed with typically very high \SNR{}. Additionally, our spectra are defined over the restricted wavelength range $4,500\leq\lambda\leq7,000$ and for our measurement of \SNR{} we need the spectrum to be accurately smoothed in the region around the feature we use to measure \SNR{}, rather than globally across the entire spectrum (although the smoothing is done globally and the \signal{} is extracted for the whole wavelength range).

We then manually review every spectrum to select a smoothing factor that optimally extracts an approximation of the noiseless spectrum, which we call the \signal{}. This smoothing factor varies per spectrum due to differences in the \SNR{} of the original spectrum. The standard deviation of the Gaussian filter we used has a range of $10\text{ \AA}\leq\sigma\leq50\text{ \AA}$ or $1.1\text{ px}<\sigma<5.8\text{ px}$ at $\R = 738$ and \spl{6355}, the location of the \lname{Si}{ii} line which is generally easily detectable in SNe Ia and plainly visible in \autoref{fig:example_specs}. The \noise{} spectrum is then obtained by simply subtracting this \signal{} from the original standardized spectrum.

\begin{figure*}[t!]
    \centering
    \includegraphics[width=0.8\linewidth]{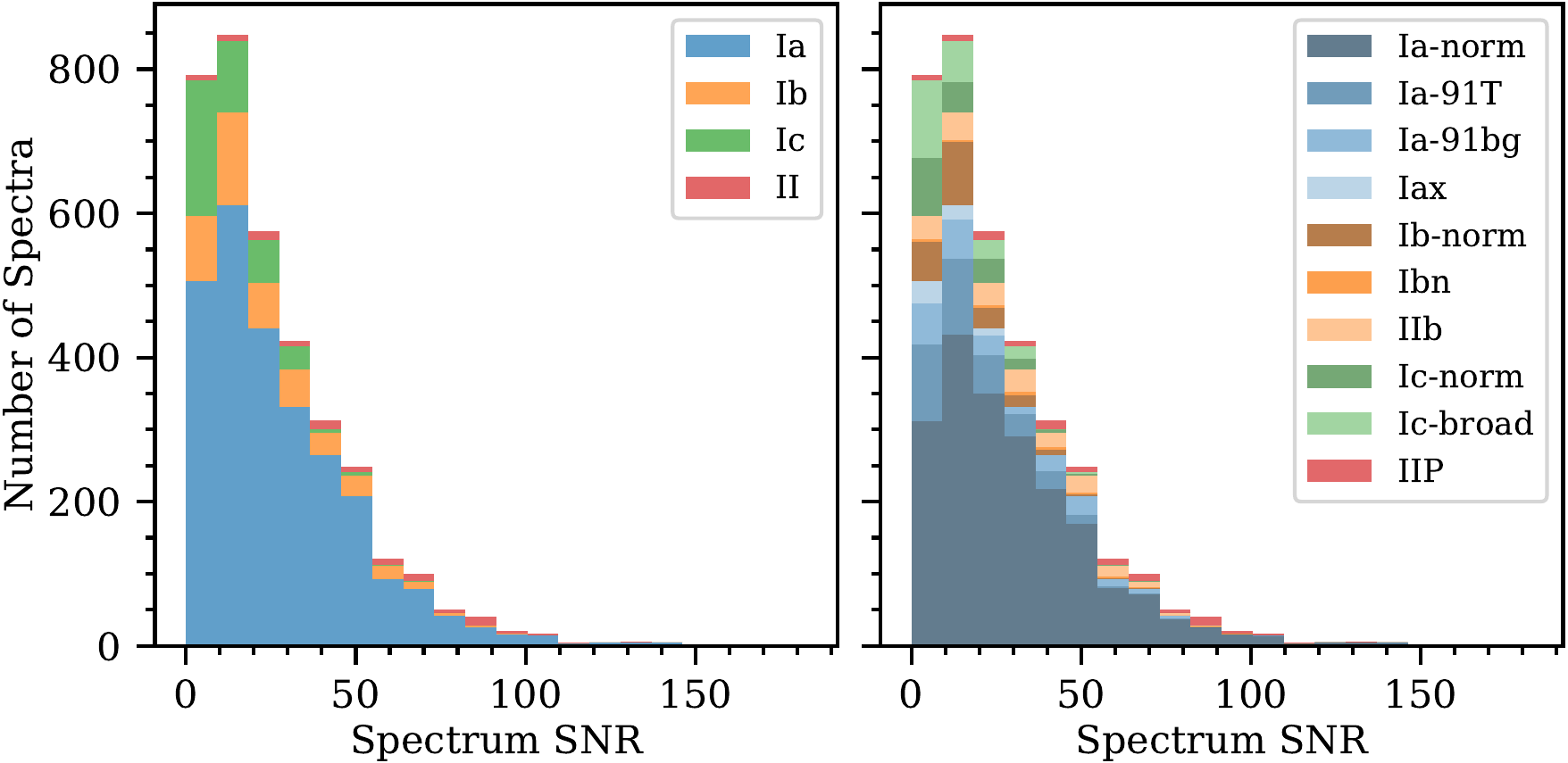}
    \caption{A stacked histogram of spectrum \SNR{} is shown for \nSpecfinal{} spectra. The median \SNR{} is $\approx20$. Left: the spectra are stacked based on their main type: Ia, Ib, Ic, II. Right: the spectra are stacked based on their subtype.}
    \label{fig:snr_histogram}
\end{figure*}

With the \signal{} in hand, we next identify a single, prominent spectral feature whose location is known \textit{a~priori} (see \autoref{sec:features} and \autoref{tab:features}). For each spectrum, the algorithm searches the predetermined wavelength region of \signal\ (\eg{} around the \lname{Si}{ii} feature for \Ias), for a local minimum --- or a local maximum in the case of \eg{} \Ibns{} at \lname{He}{i} $\lambda5876$ (see \autoref{tab:features}). The algorithm then locates the shoulders of the feature by finding the nearest local maxima (minima for emission lines) on either side of the minimum (maximum), denoted as $S_{m}$, which we also manually review. These shoulders define the boundaries of the spectral feature. A linear pseudo-continuum is constructed by connecting the two shoulders. The \textit{signal area} (in units of flux $\times$ wavelength) is then calculated via numerical integration of the area between the pseudo-continuum and the signal. We divide the \textit{signal area} by the width of the spectral feature (from shoulder to shoulder) to obtain a quantity with units of flux, which we define as $S$, the signal in our 
\SNR\ calculation as in \autoref{eq:s}. Note that in some cases this method will inevitably be impacted by line blending. Consider SNe 1993j and 2000h (type IIb) or 2005cs (type II) in \autoref{fig:features}: in each case, the line is likely blended with another blue-ward feature and a reasonable solution would have been to identify the left edge of the feature at $\lambda\approx6,200$ ($\lambda\approx6,350$ for SN 2005c) instead of $\lambda\approx5,750$. However, for the purpose of our work, what matters most is consistency. That the method consistently identifies the same region for these SN types. In addition, while we add signal $S$ by including the shallow region on the blue side of the feature, we also penalize $S$ by the overall width, such that the measured \SNR{} in both cases does not differ substantially.

To measure the noise $N$, we compute the standard deviation of the data points in the \noise\ array at $\lambda<\lambda_{S_{m}}-100\text{ \AA}$ to the left of the left shoulder and $\lambda>\lambda_{S_{m}}+100\text{ \AA}$ to the right of the right shoulder. The \SNR{} for each spectrum is then calculated as $S/N$.

While we aspired to create an automated and systematic algorithm for the calculation of \SNR, the resulting procedure required extensive manual oversight. As we reviewed each spectrum, we identified and removed unsuitable or incomplete spectra (\eg{} those with missing wavelength coverage or severe artifacts). The original dataset of $3764$ spectra was thus culled to \nSpec. \autoref{tab:spectra} documents the number of spectra removed per subtype.

\subsection{Measuring and Modifying \SNR{}}\label{sec:measuringSNR}
The distribution of original \SNR\ for the objects in our final dataset is shown in \autoref{fig:snr_histogram}. The median \SNR{} value is $\approx20$, but we note a long high \SNR{} tail. However, we caution the reader that at high \SNR{} values, our method may be impacted by small variations in the limited region where we measure the noise, losing some reliability (in practice, it is hard to distinguish by eye differences in $\SNR>100$ or $\SNR<1$, although we see impacts on the model performance below the latter threshold).

Having measured $S$ and obtained the signal array for each spectrum, we can now simulate observation at an arbitrary \SNR . For a desired target \SNR , \SNRnew , we first calculate the required new noise level: $N_{\text{new}}=S/\text{SNR}_{\text{new}}$. We then generate an array of Gaussian noise with zero mean and standard deviation $N_{\text{new}}$ and add it to the signal \S\ spectrum\footnote{We considered using the original \texttt{noise} array and amplifying it according to the desired \SNR, but doing so would have made our method susceptible to any signal information present in \texttt{noise} due to imperfect \texttt{signal} and \texttt{noise} separation, thus causing a data leakage that would impact our machine learning results.}. This yields a synthetic spectrum with the same intrinsic signal at a different \SNR .

We validate our result by re-measuring the \SNR\ of these synthetic spectra to confirm they match our desired values following the algorithm described above, but with two critical adjustments to eliminate the need for further manual review. First, the smoothing parameter for extracting the signal from the noisy synthetic spectrum is optimized by maximizing the Pearson correlation coefficient $r$ \citep{benesty2009pearson} between the \signal\ and the signal newly extracted from the \SNRnew\ spectrum. Second, we fix the locations of the spectral feature shoulders to those identified on the original spectrum. These adjustments ensure a consistent, repeatable and automated \SNR\ measurement for every synthetic spectrum.

\begin{figure*}[!hp]
    \centering
    \includegraphics[width=.8\linewidth]{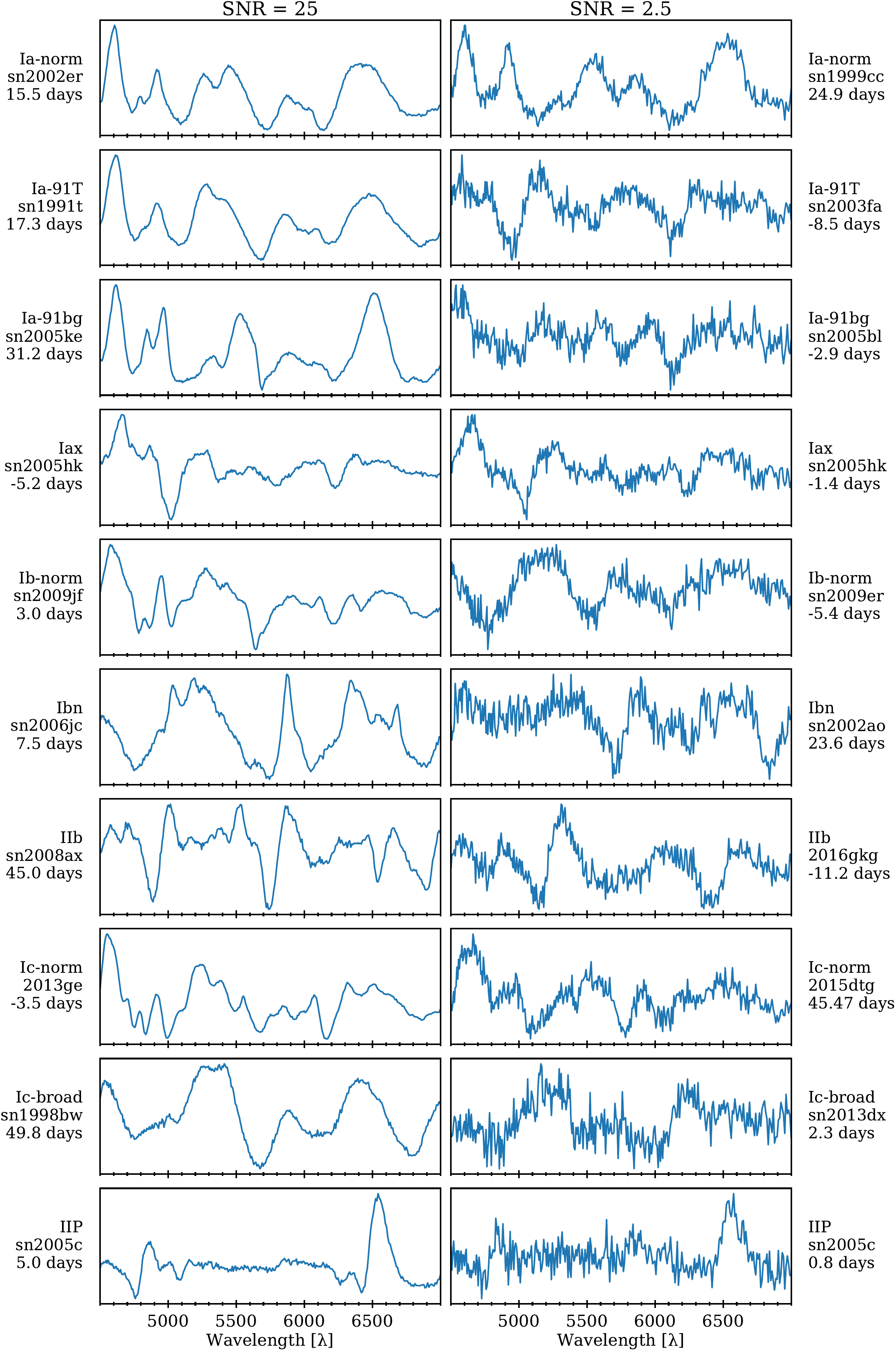}
    \caption{Example spectra are plotted for each subtype. The left and right columns show spectra with an original \SNR{} of $\approx25$ and $\approx 2.5$, respectively.}
    \label{fig:snr_hi_lo}
\end{figure*}

\subsection{Construction of the Full Dataset}
\begin{figure*}[t]
    \centering
    \includegraphics[width=1\linewidth]{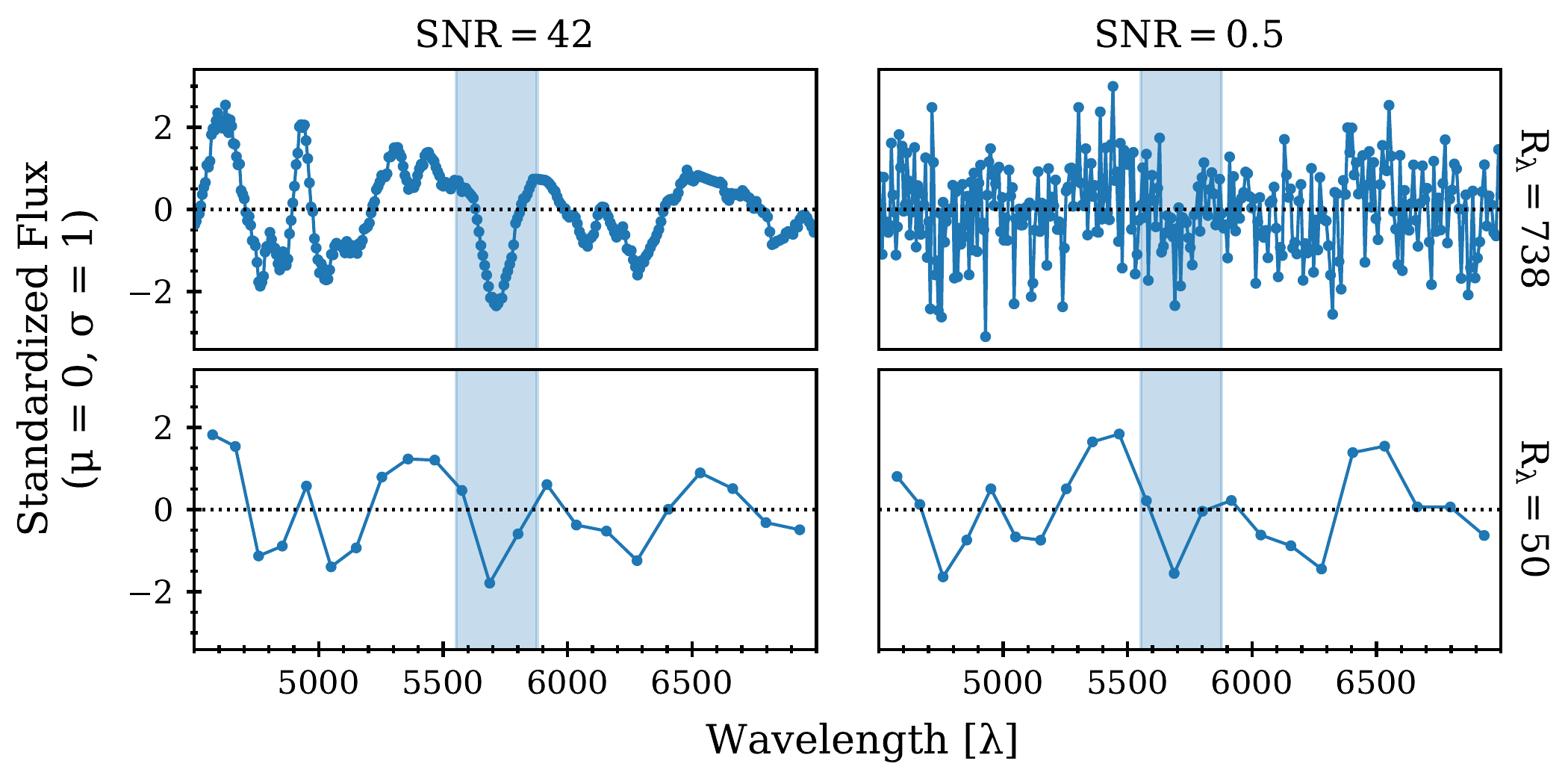}
    \caption{Supernova 1990b (Type Ic-norm, same as in \autoref{fig:example_specs}) at 4 days after peak brightness is plotted at four combinations of \SNR{} and \R: original \SNR{} of the spectrum (\SNR=$53$) and imposed $\SNRnew=0.5$, at both the original resolution $\R=738$ and $\R=50$. For $\R=50$, the spectra are linearly interpolated to a wider set of bins, simulating what a low-resolution spectrograph would observe. However, we do not train on the interpolated spectra (see \autoref{sec:r}). Additionally, as discussed in \autoref{sec:measuringSNR}, the method we developed to enable automation of \SNR{} measurements relies on specific wavelength regions for each feature as identified at $\R=738$, thus at very low $\R$ our assessment of \SNR{} may be less accurate. Thus, at low \R{} the spectra become, by eye, very similar at different \SNR{} despite the performance of the classifier being strongly impacted by both parameters (see \autoref{sec:conclusion}).}
    \label{fig:snr_r}
\end{figure*}

With this method, we generate the entire cleaned dataset of \nSpecfinal{} spectra at \numSNR{} different \SNR{} values: \SNR=[100, 50, 25, 20, 5, 2.5, 2, 1.75, 1.5, 1.25, 1, 0.9, 0.8, 0.7, 0.6, 0.5]. Following the process described in \autoref{sec:r} to lower the resolution, we then generate each of these \numSNR{} \SNR{} datasets at \numR{} different spectral resolutions: \R=[700, 600, 500, 400, 300, 200, 100, 50, 45, 40, 35, 30, 25, 20, 15, 10]. We perform a two-fold cross-validation split, doubling the total number of datasets we generate. An example spectrum is plotted in \autoref{fig:snr_r} at two different \SNR{} and \R{} values. Lastly, we also create datasets at each \R{} but with untouched \SNR{} in order to verify the performance of \abcsn{} remains consistent with the results in \citetalias{fortino_abc-sn_2026}. Altogether, we created \numTotalModels{} datasets.

As a final step, we further curate the data by identifying a set of spectra for which the re-measured \SNR\ deviated significantly from the target value, leaving a grand total of $\nSpecfinal$ spectra (see \autoref{tab:spectra}).

\subsection{Re-Training ABC-SN}\label{sec:retrain}
For each of our \numTotalModels{} versions of the original dataset, we then retrained the \texttt{ABC-SN} classifier. We made two minor but necessary adjustments to the training scheme described in \citetalias{fortino_abc-sn_2026} for this work. 

Recall that the original preprocessing in \citetalias{fortino_abc-sn_2026} defined spectra from $2,500\text{ \AA}$ – $10,000\text{ \AA}$, with non-zero data only from $4,500\text{ \AA}$ – $7,000\text{ \AA}$. We removed the zero-padding region from the input spectra. The $4,500\text{ \AA}$ – $7,000\text{ \AA}$ wavelength range contains a number of lines known to be important for classification, \eg{} the Balmer lines and the optical He lines.  It is, however, somewhat restricted. The $7,000\text{ \AA}$ – $10,000\text{ \AA}$ range accessible with several large telescopes (\eg Keck and Lick observatories) includes two very strong line complexes: the  \lname{O} \spl{7774} and the  NIR \lname{Ca} triplet. The restricted wavelength range has been chosen as the range where most objects in the \snid{} template sample have coverage. In future work, we plan to extend the capabilities of \abcsn{} to a larger and more flexible wavelength range. But for the purpose of this study, we remain consistent with the choices made in \abcsn.

We also modified the data augmentation scheme used to balance the dataset. In \citetalias{fortino_abc-sn_2026} we augmented the dataset by creating copies of spectra with the following modifications: we randomly shift the spectra by up to 225 Å on the blue end and 350 Å on the red end to simulate inaccurate redshift corrections and inject 0–4 spikes to loosely simulate telluric lines, telluric line inaccurate correction, and possible instrumental effects; in \citetalias{fortino_abc-sn_2026}, class balance was achieved solely by generating augmented copies of spectra while omitting the original copies. In this work, we ensure that both the original and augmented copies of each spectrum are included during training. In order to maintain balanced classes, this leads to more augmented copies being created than in \citetalias{fortino_abc-sn_2026} (see \autoref{tab:spectra}). The impact of this change on performance should be minor, but to confirm the published performance of \texttt{ABC-SN}, we have re-trained it on the dataset at the original \SNR{} and at all \numR{} \R{} values. Its (unchanged) performance is included in \autoref{fig:heatmap_full}. In order to minimize the performance variance induced by the small test set, as in \citetalias{fortino_abc-sn_2026}, we performed a 2-fold cross-validation split by swapping the training and test sets. As discussed in \citetalias{fortino_abc-sn_2026}, the significant imbalance combined with the requirement to keep each SN only in the training or the test set to avoid data leakage prevents finer splits.

Throughout, we will use the following metrics of performance. For each class we define:

\begin{align*}
    \textrm{Precision} &= \frac{TP}{TP + FP} \\
    \textrm{Recall} &= \frac{TP}{TP + FN} \\
    F_1 &= \frac{2TP}{2TP + FP + FN},
\end{align*}

\noindent
where $TP$ stands for True Positive, $FP$ stands for False Positive, and $FN$ stands for False Negative (see \citetalias{fortino_abc-sn_2026} for more details). Importantly, our preferred metric is the `macro $F1$-score', which is obtained by calculating the per-class $F1$-score and taking their {\it unweighted} average.  Since we balanced the training but not the test set, this is our preferred global metric of performance.

\section{Results and Discussion}\label{sec:results}
\begin{figure*}[!hp]
    \centering
    \includegraphics[width=0.8\linewidth]{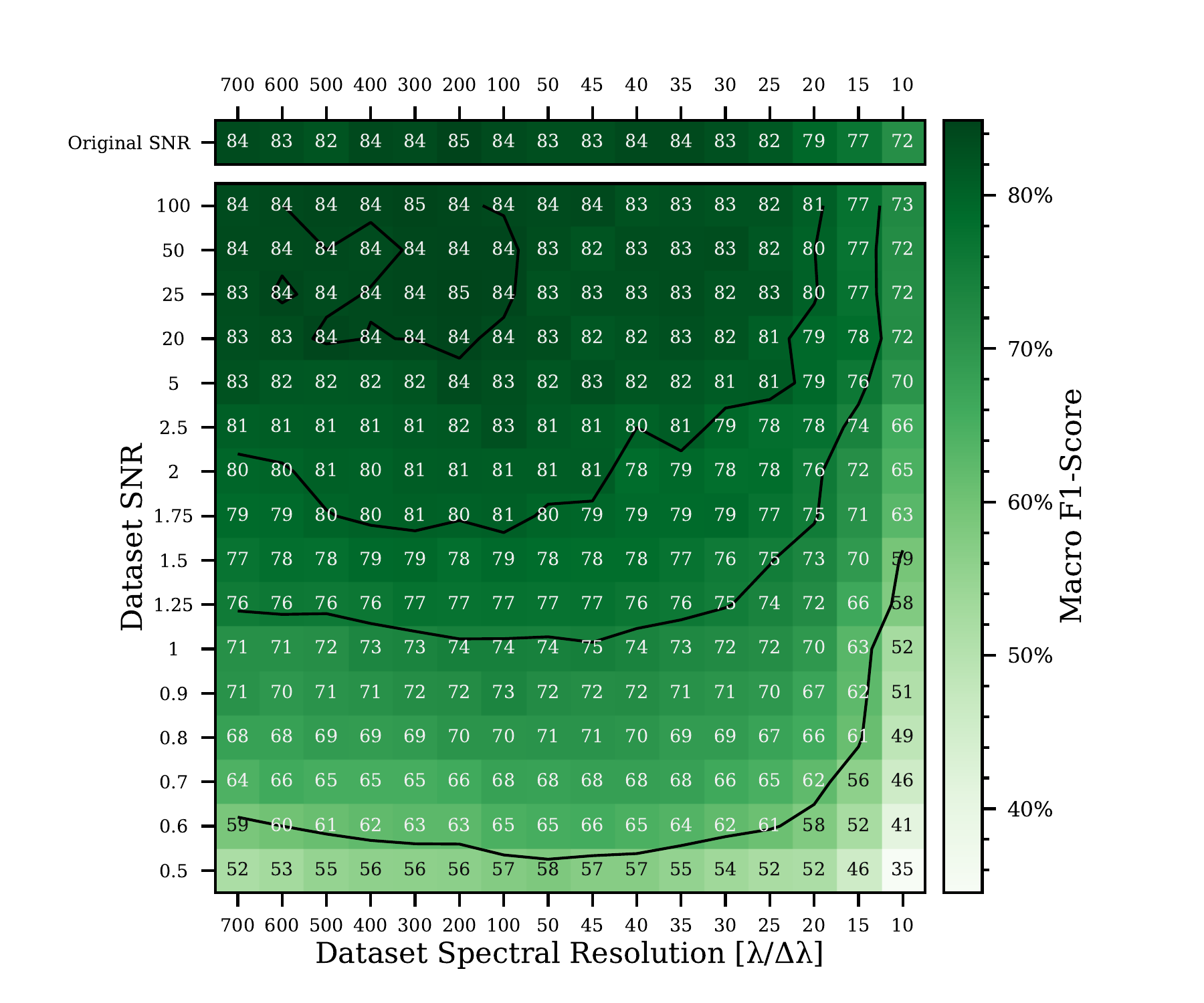}
    \caption{The macro $F1-$score of \abcsn{} is plotted on the color-axis while the \SNR{} and \R{} of the dataset it was trained on are shown on the vertical and horizontal axes. We perform a 2-fold cross-validation, so each cell corresponds to the average macro $F1-$score of two \abcsn{} models. The top row shows \abcsn{} performance when the \SNR{} of the dataset isn't changed at all. This represents a validation and extension of model performance in comparison to \citetalias{fortino_abc-sn_2026}, which was trained at $\R=100$. The median \SNR{} of the original dataset is $\approx20$. Contour lines are drawn at $60\%$, $75\%$, $80\%$, and $84\%$. \SNR{} does have a noticeable effect on model performance with small losses at $\SNR = 2$ and significant losses beginning at $\SNR{}\approx1$. Lowering \R{} has the expected effect of reducing model performance starting at $\R\approx35$. \SNR{} and \R{} have mostly independent effects on model performance, but at very low \SNR{}, the performance of \abcsn{} begins to degrade at somewhat higher \R{}.}
    \label{fig:heatmap_full}
\end{figure*}

\begin{figure}
    \centering
    \includegraphics[width=1\linewidth]{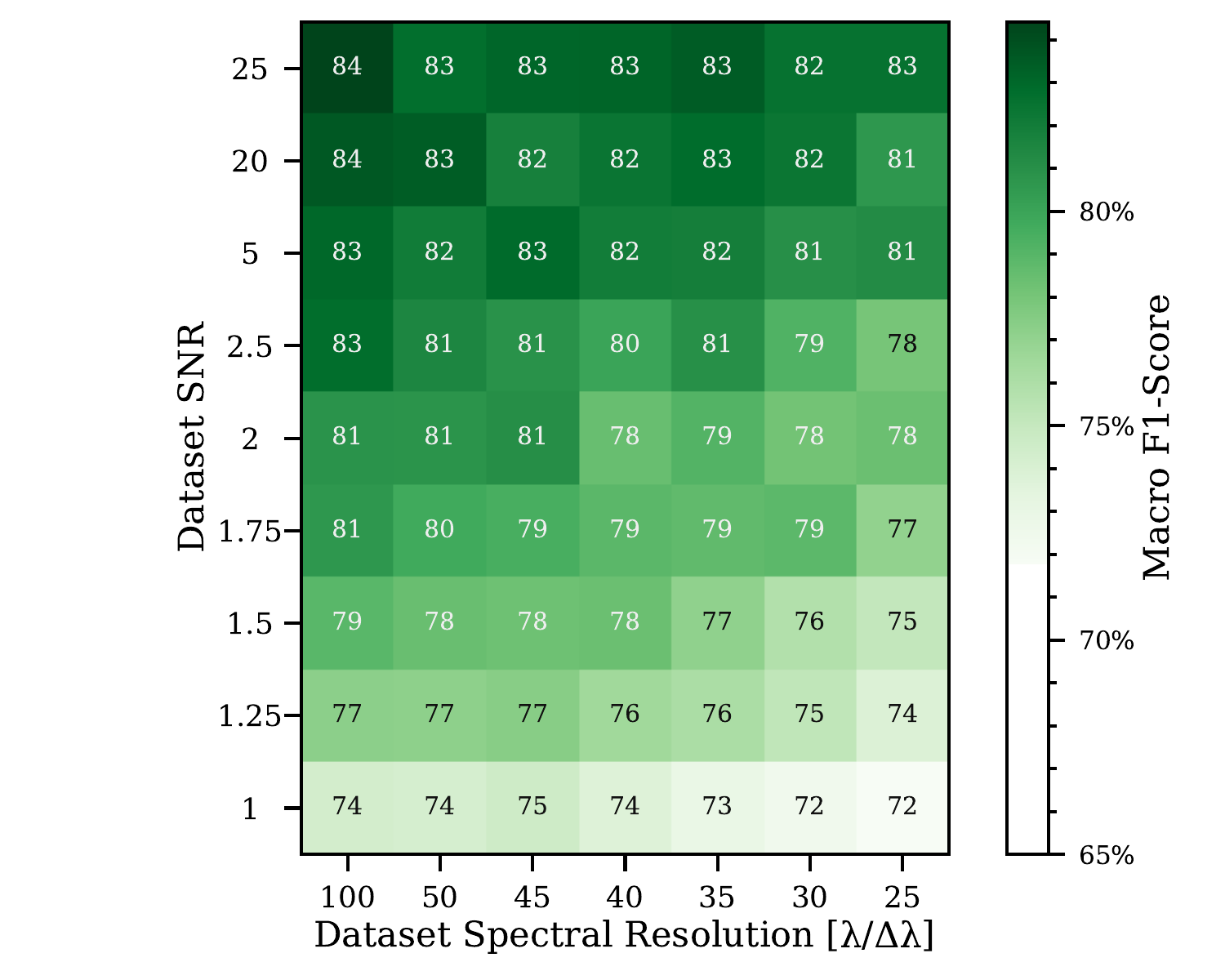}
    \caption{An abridged version of \autoref{fig:heatmap_full} is shown with a narrower dynamic range. This allows the trends of performance in the critical area of \SNR{} and \R{} to be more readily examined.}
    \label{fig:heatmap_specific}
\end{figure}

\abcsn{} was re-trained on a total of \numTotalModels{} datasets spanning \numSNR{} \SNR{} and \numR{} \R{} values. In \autoref{fig:heatmap_full}, we present test set performance metrics for the \numTotalModels{} models. Each cell represents the macro $F1$-score of \abcsn{} trained on a dataset at the corresponding \SNR{} and \R{}, averaged across the 2-fold cross-validation split. In \autoref{fig:heatmap_specific}, we zoom into the region where performance degradation starts: $100\leq\R\leq25$ and $1\leq\SNR\leq25$.

Model performance is heavily impacted by both \SNR{} and \R{}.

{$\mathbf \R$}: Classification power declines below $\R=50$ with a notable and progressive loss of performance at $20$, $15$, and $10$.
Large performance losses are expected at such low resolutions, as even the broad features are washed out. In fact, as in \citetalias{fortino_abc-sn_2026}, we find that peak performance is obtained at a resolution of $\R{}\sim300$, rather than at the highest resolution of $\R{}\sim700$. This is easily explainable by the noise reduction effects of smoothing. 

{\bf \SNR}: As a rough baseline, consider the row in \autoref{fig:heatmap_full} corresponding to $\SNR=20$, which is the approximate median \SNR{} of the original dataset. We first note that increasing the \SNR{} beyond $20$ does not help model performance. In fact, we identify $\SNR=2.5$ to show the first noticeable impact on performance. Below $\SNR=1.25$, performance erodes rapidly at all resolutions.\footnote{We remind the reader that we have a specific definition of \SNR{}, see \autoref{sec:snr}.}

\section{Conclusion and Future Work} \label{sec:conclusion}
We have conducted a systematic investigation of the minimum spectral resolution, \R{}, and signal-to-noise ratio (\SNR{}) required for the classification of SN subtypes using \abcsn{}, a state-of-the-art classifier for SN spectra recently released by our group \citepalias{fortino_abc-sn_2026}. This work was motivated by the impending plethora of data from the \lsst{} and the Argus Array, which will discover millions of SNe and place unprecedented demands on spectroscopic follow-up resources. To our knowledge, while the performance of classifiers at $\R=100$ resolution is demonstrated in practice by the \sedm{} \citep{2012SPIE.8446E..86B,blagorodnova_sed_2018,fremling_sniascore_2021}, this is the first comprehensive study to systematically evaluate classification performance as a function of both \SNR{} and \R{}.

We developed a novel, semi-automated method for measuring and modifying the \SNR{} of SN spectra (\autoref{sec:snr}). Our method is designed for situations where neither uncertainty arrays nor the original raw spectra are available. Our method smooths each spectrum with a Gaussian filter to extract the signal and noise components of the spectrum. We then measure the area of the spectral feature as in \citet{folatelli_spectral_2004} to quantify the signal, $S$, but divide the area by the feature width to obtain a unitless quantity. The process required manual review of each spectrum, and \nSpecCulled{} spectra were removed during this process, and an additional \nSpecClipped{} in the next step, as their modified \SNR{} resulted in inconsistent target values, indicating inaccuracies in the first \SNR{} evaluation.

We created a catalog of \numTotalModels{} datasets at various combinations of \SNR{} (from $100$ to $0.5$, as well as unchanged) and \R{} (from $500$ to $10$). Our data include most of the common SN subtypes: SN Ia, Ia-91T, Ia-91bg, Iax, Ib, Ic, broad-lined Ic, IIb, IIP, and Ibn. We re-trained \abcsn{} on each dataset and evaluated the macro $F1-$score, which appropriately accounts for the severe class imbalance in our test set, in a 2-fold cross-validation framework ---the number of folds limited by the requirement to maintain class balance and simultaneously avoid data leakage (\autoref{sec:retrain}).

We found that classification performance peaks at a spectral resolutions $\R=\sim200$ and $\SNR\sim25$. Performance is only minimally reduced between $\R=50$ and $\R=25$, well below the resolution of \sedm{}, but declines rapidly below $\R=20$. We also found that increasing the \SNR{} of the dataset above $5$ did not seem to impact model performance significantly. For comparison, under our definition, the median \SNR{} of the original dataset is $\approx20$. 

Our results demonstrate that classification of SN spectra into a refined taxonomy that separates, for example, between different subtypes of SESNe, is possible at low resolution and low \SNR{} with no loss in model performance down to $\R=50$ and $\SNR=5$. 

These thresholds provide guidance for observers and designers of future instruments and small observatories, as well as empowering the community to take better advantage of their existing spectroscopic resources in order to maximize the science returns of large discovery surveys. This result is intended to support and encourage a responsible use of spectroscopic resources in an era that started with ZTF and culminates with Rubin \lsst{} and Argus Array, where the rate of transient discovery will vastly outmatch the capacity of current spectroscopic facilities. Yet if observers can take advantage of low-resolution spectrographs and/or reduce their exposure time of spectra collected for classification, this will make more resources available for the targeted collection of high-quality spectra of selected transients. This result could also expand access to LSST science (and perhaps data via in-kind contributions) to communities that may not have access to high-resolution spectrographs on large telescopes, with which high-quality spectra can be obtained, but that may have access to smaller telescopes equipped with spectrographs where observations at simultaneously high \SNR{} and \R{} are unfeasible for targets within the reach of \emph{LSST}, but lowering either one of these parameters would allow them to reach into the \emph{LSST} discoveries stack, and may offer information to direct their investments into building relatively inexpensive instrumentation for spectroscopic classification.

A significant shortcoming of this result is that the functional definition of $\SNR{}$ we developed and adopted cannot be inferred trivially before collecting a spectrum of the target object, such that these results cannot be used directly to choose, for example, integration time when planning the collection of a new spectrum. Rather, our results should be used as guidance: adopting a two-stage strategy, where the initial spectra are collected at both low \R{} and low \SNR{}, leveraging small observatories, using short integration time, and planning for appropriate \SNR{} on rebinned spectra instead of native resolution ones, would support classification and identification of interesting targets while preserving scarce spectroscopic resources (scarce compared to the flow of discoveries in the LSST era). While this strategy is not in and of itself novel (it is, after all, the thrust behind the creation of the \sedm{}), this study is, to our knowledge, the first systematic quantification of the impacts of \R{} and \SNR{} on classification.

We note that, for simplicity and clarity, these results rely on models trained on transients with uniform \SNR{} and resolution, while a training dataset would generally vary in \SNR{}. However, the performance of \abcsn{} trained on the original spectra is consistent with the performance of the model trained on $\SNR{}=20$, the median \SNR{} value of the original dataset. We expect this result to hold with models trained on a dataset with a range of \SNR{}, but warn the user about systematic differences in \SNR{} for different subtypes that may bias a model result.

Finding more spectra of the least numerous classes would go far towards increasing the confidence in \abcsn{}'s performance on those classes. Additionally, the scope of this analysis could be broadened by expanding the dataset with a wider variety of transients. In future work, we will expand the capabilities of \abcsn{} to a broader set of transients and simpler preprocessing steps, building robustness into the model to contamination from galaxy lines and continuum, and supporting a broader and more flexible wavelength range. These and other aspects of the data and practices used in training are important drivers of model performance and our results should be checked against this improved version of \abcsn{} as well as other classifiers, but there is no principle reason to think that what we found in our work would not generalize to other models. For now, we want to encourage systematic studies of minimal requirements to enable scientific return by identifying and addressing infrastructure bottlenecks.

This work is reproducible. All material that produced these results is available at a dedicated GitHub repository,\footnote{\url{https://github.com/FoxFortino/spec_res}} which includes notebooks that reproduce the figures in this work.

\begin{acknowledgments}
FBB is supported in part by NSF AST Awards Number 2511639 and 2308016. WFF is supported in part by NSF HDR Award Number 2123264. FBB thanks long-time friend and colleague Lou Strolger. The idea for this work arose in discussions with Lou. 

M.M. acknowledge support in part from ADAP program grant No. 80NSSC22K0486, from the NSF grant AST-2206657 and from the National Science Foundation under Cooperative Agreement 2421782 and the Simons Foundation grant MPS-AI-00010515 awarded to the NSF-Simons AI Institute for Cosmic Origins — CosmicAI, https://www.cosmicai.org/.
\end{acknowledgments}

This work made use of the following software packages: 

\texttt{NumPy} \citep{harris2020array};

\texttt{Pandas}\citep{mckinney2010data};

\texttt{Matplotlib} \citep{4160265};

\texttt{SciPy} \citep{virtanen2020scipy};

\texttt{Astropy} \citep{robitaille2013astropy};

\texttt{scikit-learn} \citep{scikit-learn};

\texttt{TensorFlow} \citep{tensorflow2015-whitepaper};

\texttt{Keras} \citep{chollet2015keras};

\texttt{KerasHub} \citep{kerashub2024};


\texttt{tqdm} \citep{da2019tqdm}.

\begin{appendix}\label{sec:appendix}
Below we list all SN spectra used in our work, including those removed at each step of preprocessing: \autoref{tab:culled_spectra}, and \autoref{tab:clipped_spectra}. 

\begin{longtable}{ccc|ccc|ccc}
    \toprule
    Name & Phase (days) & Subtype & Name & Phase (days) & Subtype & Name & Phase (days) & Subtype \\
    \midrule
    sn1981b  &  26.3  &  Ia-norm  & sn2005cg  &  6.7  &  Ia-norm  & 2010as  &  -9.17  &  IIb \\
sn1981b  &  33.2  &  Ia-norm  & sn2006al  &  2.5  &  Ia-norm  & 2010as  &  -9.17  &  IIb \\
sn1984a  &  -7.0  &  Ia-norm  & sn2006al  &  3.4  &  Ia-norm  & 2010as  &  -8.17  &  IIb \\
sn1984a  &  -6.7  &  Ia-norm  & sn2006al  &  4.3  &  Ia-norm  & 2010as  &  -8.17  &  IIb \\
sn1984a  &  -5.9  &  Ia-norm  & sn2006al  &  5.4  &  Ia-norm  & 2010as  &  -6.17  &  IIb \\
sn1984a  &  -4.9  &  Ia-norm  & sn2006al  &  6.3  &  Ia-norm  & 2010as  &  -6.17  &  IIb \\
sn1984a  &  -2.8  &  Ia-norm  & sn2006al  &  9.1  &  Ia-norm  & 2010as  &  -5.17  &  IIb \\
sn1984a  &  8.1  &  Ia-norm  & sn2006cj  &  -2.1  &  Ia-norm  & 2010as  &  -5.17  &  IIb \\
sn1984a  &  17.0  &  Ia-norm  & sn2006cj  &  -1.2  &  Ia-norm  & 2011hs  &  -8.17  &  IIb \\
sn1984a  &  19.1  &  Ia-norm  & sn2006cj  &  -0.2  &  Ia-norm  & 2011hs  &  -5.17  &  IIb \\
sn1984a  &  26.2  &  Ia-norm  & sn2006cj  &  4.4  &  Ia-norm  & 2011hs  &  27.83  &  IIb \\
sn1989b  &  2.6  &  Ia-norm  & sn2006cj  &  5.4  &  Ia-norm  & 2016gkg  &  -18.16  &  IIb \\
sn1989b  &  3.5  &  Ia-norm  & sn2006cj  &  8.2  &  Ia-norm  & 2016gkg  &  -18.12  &  IIb \\
sn1989b  &  4.6  &  Ia-norm  & sn2006cj  &  9.1  &  Ia-norm  & 2016gkg  &  -17.15  &  IIb \\
sn1989b  &  5.5  &  Ia-norm  & sn2006cj  &  11.0  &  Ia-norm  & LSQ2014efd  &  -12.05  &  Ic-norm \\
sn1989b  &  6.6  &  Ia-norm  & sn2007ae  &  3.3  &  Ia-norm  & LSQ2014efd*  &  31.93  &  Ic-norm \\
sn1989b  &  7.5  &  Ia-norm  & sn1991t  &  -12.8  &  Ia-91T  & LSQ2014efd*  &  31.93  &  Ic-norm \\
sn1989b  &  8.6  &  Ia-norm  & sn1991t  &  -11.5  &  Ia-91T  & sn1983v  &  -9.0  &  Ic-norm \\
sn1989b  &  9.5  &  Ia-norm  & sn1998ab  &  6.7  &  Ia-91T  & sn1983v  &  -4.0  &  Ic-norm \\
sn1989b  &  10.6  &  Ia-norm  & sn1998e  &  -9.4  &  Ia-91T  & sn1983v  &  4.0  &  Ic-norm \\
sn1989b  &  11.5  &  Ia-norm  & sn2005hj  &  3.3  &  Ia-91T  & sn1983v  &  11.9  &  Ic-norm \\
sn1989b  &  12.6  &  Ia-norm  & sn2005hj  &  9.9  &  Ia-91T  & sn1983v  &  12.9  &  Ic-norm \\
sn1989b  &  13.4  &  Ia-norm  & sn2005hj  &  17.4  &  Ia-91T  & sn1983v  &  13.9  &  Ic-norm \\
sn1990n  &  1.6  &  Ia-norm  & sn2005hj  &  24.0  &  Ia-91T  & sn1983v  &  14.9  &  Ic-norm \\
sn1990o  &  18.6  &  Ia-norm  & sn2005hj  &  32.5  &  Ia-91T  & sn1983v  &  15.9  &  Ic-norm \\
sn1990o  &  19.4  &  Ia-norm  & sn1986g  &  0.3  &  Ia-91bg  & sn1983v  &  16.9  &  Ic-norm \\
sn1992a  &  6.9  &  Ia-norm  & sn1986g  &  1.3  &  Ia-91bg  & sn1983v  &  17.9  &  Ic-norm \\
sn1992a  &  15.8  &  Ia-norm  & sn1986g  &  20.7  &  Ia-91bg  & sn1992ar  &  3.4  &  Ic-norm \\
sn1992a  &  23.8  &  Ia-norm  & sn1986g  &  21.7  &  Ia-91bg  & sn1994i  &  33.06  &  Ic-norm \\
sn1992a  &  27.8  &  Ia-norm  & sn1986g  &  22.7  &  Ia-91bg  & sn1994i  &  34.06  &  Ic-norm \\
sn1994d  &  26.2  &  Ia-norm  & sn1986g  &  31.7  &  Ia-91bg  & sn2004fe*  &  -6.78  &  Ic-norm \\
sn1994d  &  48.0  &  Ia-norm  & sn1986g  &  41.2  &  Ia-91bg  & sn2011bm  &  30.3  &  Ic-norm \\
sn1994m  &  39.8  &  Ia-norm  & sn1998de  &  -7.2  &  Ia-91bg  & 2013ge  &  -14.5  &  Ic-norm \\
sn1994q  &  18.6  &  Ia-norm  & sn1999by  &  -5.0  &  Ia-91bg  & 2013ge  &  -4.5  &  Ic-norm \\
sn1994s  &  -3.0  &  Ia-norm  & sn1999by  &  -4.8  &  Ia-91bg  & 2013ge  &  5.5  &  Ic-norm \\
sn1994t  &  0.8  &  Ia-norm  & sn2005hk  &  -4.2  &  Iax  & 2013ge  &  14.5  &  Ic-norm \\
sn1995d  &  3.6  &  Ia-norm  & sn2005hk  &  -3.4  &  Iax  & 2013ge  &  18.5  &  Ic-norm \\
sn1995d  &  4.2  &  Ia-norm  & iPTF2013bvn  &  -0.8  &  Ib-norm  & 2013ge  &  37.5  &  Ic-norm \\
sn1995e  &  -2.9  &  Ia-norm  & sn1984l  &  8.0  &  Ib-norm  & 2013ge  &  45.5  &  Ic-norm \\
sn1995e  &  -0.9  &  Ia-norm  & sn1984l  &  9.0  &  Ib-norm  & 2017ein  &  15.22  &  Ic-norm \\
sn1995e  &  1.0  &  Ia-norm  & sn1984l  &  11.9  &  Ib-norm  & sn1998bw  &  -11.2  &  Ic-broad \\
sn1995e  &  3.9  &  Ia-norm  & sn1984l  &  12.9  &  Ib-norm  & sn2003dh  &  -9.8  &  Ic-broad \\
sn1995e  &  5.9  &  Ia-norm  & sn1984l  &  27.9  &  Ib-norm  & sn2003dh  &  -7.3  &  Ic-broad \\
sn1996ab  &  1.3  &  Ia-norm  & sn2005bf  &  -1.9  &  Ib-norm  & sn2003dh  &  12.4  &  Ic-broad \\
sn1996x  &  11.9  &  Ia-norm  & sn2005hg  &  22.71  &  Ib-norm  & sn2003dh  &  14.1  &  Ic-broad \\
sn1998bu  &  1.2  &  Ia-norm  & sn2008d  &  -12.8  &  Ib-norm  & sn2006aj  &  1.8  &  Ic-broad \\
sn1998co  &  28.4  &  Ia-norm  & sn2009jf  &  -10.0  &  Ib-norm  & sn2009bb  &  41.5  &  Ic-broad \\
sn1998dx  &  -1.7  &  Ia-norm  & sn2009jf  &  -9.0  &  Ib-norm  & sn2009nz  &  0.0  &  Ic-broad \\
sn1998dx  &  -0.7  &  Ia-norm  & 2012au  &  21.42  &  Ib-norm  & sn2010ma  &  -5.5  &  Ic-broad \\
sn1998dx  &  2.1  &  Ia-norm  & 2012au  &  33.42  &  Ib-norm  & sn2010ma  &  5.2  &  Ic-broad \\
sn1998ec  &  38.2  &  Ia-norm  & sn2000er  &  4.7  &  Ibn  & sn2010ma  &  18.2  &  Ic-broad \\
sn1998eg  &  16.8  &  Ia-norm  & sn1993j  &  11.0  &  IIb  & sn2012ap  &  1.0  &  Ic-broad \\
sn1998eg  &  18.8  &  Ia-norm  & sn1993j  &  12.0  &  IIb  & sn2012bz  &  3.9  &  Ic-broad \\
sn1998v  &  30.8  &  Ia-norm  & sn1993j  &  35.7  &  IIb  & sn2012bz  &  8.6  &  Ic-broad \\
sn1999cc  &  18.2  &  Ia-norm  & sn2009mg  &  2.9  &  IIb  & sn2013cq  &  -1.3  &  Ic-broad \\
sn1999cl  &  4.6  &  Ia-norm  & sn2009mg  &  2.9  &  IIb  & sn2013dx  &  -5.0  &  Ic-broad \\
sn2000fa  &  -10.4  &  Ia-norm  & sn2009mg  &  38.5  &  IIb  & sn2014ad  &  27.49  &  Ic-broad \\
sn2002er  &  -6.0  &  Ia-norm  & sn2009mg  &  38.5  &  IIb  & sn2014ad  &  37.03  &  Ic-broad \\
sn2002er  &  -5.4  &  Ia-norm  & sn2011dh  &  -9.6  &  IIb  & sn1992h  &  24.5  &  IIP \\
sn2003cg  &  -8.9  &  Ia-norm  & sn2011dh  &  37.4  &  IIb  & sn1999em  &  44.1  &  IIP \\
sn2003cg  &  -2.1  &  Ia-norm  & sn2011ei  &  -3.0  &  IIb  & sn2005c  &  -1.2  &  IIP \\
sn2005cg  &  -4.0  &  Ia-norm  & sn2013df  &  -4.42  &  IIb  & sn2005c  &  4.2  &  IIP \\
sn2005cg  &  -0.1  &  Ia-norm  & sn2013df  &  -4.42  &  IIb  &   &   & \\
sn2005cg  &  4.8  &  Ia-norm  & 2010as  &  -13.17  &  IIb  &    &   & \\
    \bottomrule
    \caption{Spectra removed from the original dataset: After manual inspection, these 190 spectra were found to have missing data between $4,500\text{ \AA} - 7,000\text{ \AA}$. Two spectra were found to have been duplicated, LSQ2014efd at $31.93 \text { days}$ and sn2004fe at $-6.78\text{ days}$, so one copy of each was removed. The LSQ2014efd spectrum was found to be too noisy, and ultimately removed. All three removed spectra are annotated with an asterisk above.}
    \label{tab:culled_spectra}
\end{longtable}

\begin{longtable}{ccc|ccc|ccc}
    \toprule
    Name & Phase (days) & Subtype &  Name & Phase (days) & Subtype &  Name & Phase (days) & Subtype \\
    \midrule

    sn1989b  &  17.5  &  Ia-norm  & sn2001eh  &  36.5  &  Ia-91T  & sn2008bo  &  -3.2  &  IIb \\
sn1994q  &  11.8  &  Ia-norm  & sn2005m  &  5.1  &  Ia-91T  & sn2008bo  &  -1.2  &  IIb \\
sn1998bp  &  24.8  &  Ia-norm  & sn2005m  &  24.7  &  Ia-91T  & sn2008bo  &  30.8  &  IIb \\
sn1998bp  &  25.8  &  Ia-norm  & sn2005m  &  40.3  &  Ia-91T  & sn2011ei  &  -4.0  &  IIb \\
sn1998bp  &  27.7  &  Ia-norm  & sn1991bg  &  15.8  &  Ia-91bg  & sn2011ei  &  8.0  &  IIb \\
sn1998bp  &  29.6  &  Ia-norm  & sn1997c  &  3.9  &  Ia-91bg  & sn2012p  &  -10.96  &  IIb \\
sn2000bk  &  26.5  &  Ia-norm  & sn1997c  &  11.9  &  Ia-91bg  & sn2012p  &  -6.96  &  IIb \\
sn2000d  &  -0.9  &  Ia-norm  & sn1997c  &  13.0  &  Ia-91bg  & sn2012p  &  1.04  &  IIb \\
sn2000dg  &  -0.8  &  Ia-norm  & sn1997c  &  19.7  &  Ia-91bg  & sn2012p  &  8.04  &  IIb \\
sn2000fa  &  37.5  &  Ia-norm  & sn1997c  &  20.8  &  Ia-91bg  & sn2012p  &  29.61  &  IIb \\
sn2002ck  &  29.3  &  Ia-norm  & sn1997c  &  21.7  &  Ia-91bg  & 2016gkg  &  -6.67  &  IIb \\
sn2002ck  &  37.1  &  Ia-norm  & sn2000c  &  20.9  &  Ia-91bg  & 2016gkg  &  -0.7  &  IIb \\
sn2002de  &  -6.1  &  Ia-norm  & sn2000c  &  26.8  &  Ia-91bg  & 2016gkg  &  1.36  &  IIb \\
sn2002de  &  8.5  &  Ia-norm  & sn2002cf  &  -5.4  &  Ia-91bg  & sn1990b  &  5.9  &  Ic-norm \\
sn2002do  &  16.7  &  Ia-norm  & sn2002e  &  9.0  &  Ia-91bg  & sn1990b  &  6.0  &  Ic-norm \\
sn2002do  &  22.6  &  Ia-norm  & sn2002e  &  11.0  &  Ia-91bg  & sn2004fe  &  22.22  &  Ic-norm \\
sn2002do  &  30.4  &  Ia-norm  & sn2002e  &  26.7  &  Ia-91bg  & sn2004ge  &  12.2  &  Ic-norm \\
sn2002hw  &  -5.2  &  Ia-norm  & sn2002fb  &  17.0  &  Ia-91bg  & sn2005az  &  17.3  &  Ic-norm \\
sn2002hw  &  17.5  &  Ia-norm  & sn2002fb  &  22.9  &  Ia-91bg  & sn2005az  &  32.3  &  Ic-norm \\
sn2002hw  &  20.4  &  Ia-norm  & sn2002fb  &  26.8  &  Ia-91bg  & sn2007gr  &  7.0  &  Ic-norm \\
sn2003cg  &  -0.2  &  Ia-norm  & sn2005bl  &  32.2  &  Ia-91bg  & sn2007gr  &  10.0  &  Ic-norm \\
sn2003cg  &  2.8  &  Ia-norm  & sn2006bz  &  -1.4  &  Ia-91bg  & sn2007gr  &  13.0  &  Ic-norm \\
sn2003cg  &  6.8  &  Ia-norm  & sn2006h  &  16.3  &  Ia-91bg  & sn2007gr  &  14.0  &  Ic-norm \\
sn2003cg  &  7.7  &  Ia-norm  & sn2007al  &  11.3  &  Ia-91bg  & sn2007gr  &  16.0  &  Ic-norm \\
sn2003ch  &  13.5  &  Ia-norm  & sn2007ba  &  2.085  &  Ia-91bg  & sn2011bm  &  -7.6  &  Ic-norm \\
sn2003ic  &  3.4  &  Ia-norm  & sn2004gv  &  13.3  &  Ib-norm  & sn2011bm  &  -0.9  &  Ic-norm \\
sn2003ic  &  8.0  &  Ia-norm  & sn2004gv  &  49.3  &  Ib-norm  & sn2011bm  &  16.4  &  Ic-norm \\
sn2003it  &  30.9  &  Ia-norm  & sn2005bf  &  31.7  &  Ib-norm  & sn2011bm  &  49.8  &  Ic-norm \\
sn2003it  &  31.9  &  Ia-norm  & sn2005hg  &  -0.29  &  Ib-norm  & 2013ge  &  -15.5  &  Ic-norm \\
sn2003iv  &  6.986  &  Ia-norm  & sn2006ep  &  45.5  &  Ib-norm  & LSQ2014efd  &  3.93  &  Ic-norm \\
sn2004dt  &  26.4  &  Ia-norm  & sn2008d  &  18.0  &  Ib-norm  & 2015dtg  &  -2.39  &  Ic-norm \\
sn2004l  &  2.5  &  Ia-norm  & sn2009jf  &  24.0  &  Ib-norm  & sn2003jd  &  26.4  &  Ic-broad \\
sn2004l  &  3.4  &  Ia-norm  & sn2009jf  &  31.0  &  Ib-norm  & PTF2010vgv  &  36.7  &  Ic-broad \\
sn2004l  &  21.8  &  Ia-norm  & sn2009jf  &  46.6  &  Ib-norm  & sn2012ap  &  14.0  &  Ic-broad \\
sn2004l  &  27.5  &  Ia-norm  & sn2000er  &  6.7  &  Ibn  & sn2013dx  &  -5.0  &  Ic-broad \\
sn2005eu  &  8.3  &  Ia-norm  & sn2000er  &  8.7  &  Ibn  & sn2013dx  &  23.0  &  Ic-broad \\
sn2005hf  &  8.2  &  Ia-norm  & sn2000er  &  9.7  &  Ibn  & sn1999em  &  27.2  &  IIP \\
sn2005hf  &  12.9  &  Ia-norm  & sn2002ao  &  30.6  &  Ibn  & sn1999em  &  32.5  &  IIP \\
sn2005hf  &  14.9  &  Ia-norm  & sn2006jc  &  15.5  &  Ibn  & sn1999em  &  34.3  &  IIP \\
sn2005hf  &  17.8  &  Ia-norm  & sn2006jc  &  17.5  &  Ibn  & sn1999em  &  35.3  &  IIP \\
sn2005kc  &  29.4  &  Ia-norm  & sn2006jc  &  18.5  &  Ibn  & sn1999em  &  37.3  &  IIP \\
sn2006bt  &  32.9  &  Ia-norm  & sn2006jc  &  19.5  &  Ibn  & sn1999em  &  39.2  &  IIP \\
sn2006cm  &  -1.9  &  Ia-norm  & sn2006jc  &  30.5  &  Ibn  & sn1999gi  &  26.9  &  IIP \\
sn2006cm  &  -1.0  &  Ia-norm  & sn2006jc  &  33.5  &  Ibn  & sn2004et  &  33.1  &  IIP \\
sn2006s  &  -2.6  &  Ia-norm  & sn2006jc  &  35.5  &  Ibn  & sn2004et  &  34.1  &  IIP \\
sn2006s  &  4.1  &  Ia-norm  & sn2006jc  &  38.5  &  Ibn  & sn2004et  &  34.5  &  IIP \\
sn1991t  &  -9.475  &  Ia-91T  & sn2006jc  &  42.5  &  Ibn  & sn2004et  &  35.1  &  IIP \\
sn1991t  &  23.5  &  Ia-91T  & sn2006jc  &  43.0  &  Ibn  & sn2004et  &  39.5  &  IIP \\
sn1995ac  &  -5.614  &  Ia-91T  & sn2006jc  &  47.5  &  Ibn  & sn2004et  &  42.0  &  IIP \\
sn1995bd  &  -12.0  &  Ia-91T  & sn1993j  &  31.6  &  IIb  & sn2004et  &  47.4  &  IIP \\
sn1995bd  &  -10.1  &  Ia-91T  & sn1996cb  &  3.6  &  IIb  & sn2005c  &  9.0  &  IIP \\
sn1997br  &  -8.0  &  Ia-91T  & sn1998fa  &  18.3  &  IIb  & sn2005c  &  9.8  &  IIP \\
sn1998ab  &  -7.8  &  Ia-91T  & sn2000h  &  -0.1  &  IIb  & sn2005c  &  10.0  &  IIP \\
sn1999ac  &  26.7  &  Ia-91T  & sn2000h  &  16.8  &  IIb  & sn2005c  &  12.7  &  IIP \\
sn2001eh  &  -2.0  &  Ia-91T  & sn2006el  &  9.82  &  IIb  & sn2005c  &  26.0  &  IIP \\
sn2001eh  &  4.7  &  Ia-91T  & sn2006el  &  10.82  &  IIb  & sn2005c  &  27.0  &  IIP \\
sn2001eh  &  5.7  &  Ia-91T  & sn2006el  &  11.82  &  IIb  & sn2005c  &  29.8  &  IIP \\
sn2001eh  &  24.9  &  Ia-91T  & sn2006el  &  12.82  &  IIb  & sn2005c  &  31.7  &  IIP \\
sn2001eh  &  28.8  &  Ia-91T  & sn2006el  &  15.82  &  IIb  & sn2005c  &  39.7  &  IIP \\
sn2001eh  &  35.5  &  Ia-91T  & sn2006el  &  16.82  &  IIb  &  &   &  \\
    \bottomrule
    \caption{Spectra removed from the original dataset: After simulating datasets at $\SNR=100$ and $\SNR=50$, \nSpecClipped{} spectra were removed that had a re-measured \SNR{} above the 95th percentile in both simulated datasets.}
    \label{tab:clipped_spectra}
\end{longtable}
\end{appendix}


\bibliography{references,additionalref}{}

@incollection{benesty2009pearson,
  title={Pearson correlation coefficient},
  author={Benesty, Jacob and Chen, Jingdong and Huang, Yiteng and Cohen, Israel},
  booktitle={Noise reduction in speech processing},
  pages={1--4},
  year={2009},
  publisher={Springer}
}

@article{horne1986optimal,
  title={An optimal extraction algorithm for CCD spectroscopy.},
  author={Horne, Keith},
  journal={Publications of the Astronomical Society of the Pacific},
  volume={98},
  number={604},
  pages={609--617},
  year={1986},
  publisher={The Astronomical Society of the Pacific}
}

@article{blondin2006using,
  title={Using line profiles to test the fraternity of type Ia supernovae at high and low redshifts},
  author={Blondin, St{\'e}phane and Dessart, Luc and Leibundgut, Bruno and Branch, David and H{\"o}flich, Peter and Tonry, John L and Matheson, Thomas and Foley, Ryan J and Chornock, Ryan and Filippenko, Alexei V and others},
  journal={The Astronomical Journal},
  volume={131},
  number={3},
  pages={1648--1666},
  year={2006}
}

@article{hoffmann2014non,
  title={Non-LTE models for synthetic spectra of Type Ia supernovae-IV. A modified Feautrier scheme for opacity-sampled pseudo-continua at high expansion velocities and application to synthetic SN Ia spectra},
  author={Hoffmann, TL and Sauer, DN and Pauldrach, AWA and Hultzsch, PJN},
  journal={Astronomy \& Astrophysics},
  volume={569},
  pages={A62},
  year={2014},
  publisher={EDP Sciences}
}

@article{Shen_2010,
doi = {10.1088/0004-637X/715/2/767},
url = {https://doi.org/10.1088/0004-637X/715/2/767},
year = {2010},
month = {may},
publisher = {The American Astronomical Society},
volume = {715},
number = {2},
pages = {767},
author = {Shen, Ken J. and Kasen, Dan and Weinberg, Nevin N. and Bildsten, Lars and Scannapieco, Evan},
title = {THERMONUCLEAR.Ia SUPERNOVAE FROM HELIUM SHELL DETONATIONS: EXPLOSION MODELS AND OBSERVABLES},
journal = {The Astrophysical Journal}
}

@article{PadillaGonzalez_2024,
doi = {10.3847/1538-4357/ad19c9},
url = {https://doi.org/10.3847/1538-4357/ad19c9},
year = {2024},
month = {mar},
publisher = {The American Astronomical Society},
volume = {964},
number = {2},
pages = {196},
author = {Padilla Gonzalez, E. and Howell, D. A. and Terreran, G. and McCully, C. and Newsome, M. and Burke, J. and Farah, J. and Pellegrino, C. and Bostroem, K. A. and Hosseinzadeh, G. and Pearson, J. and Sand, D. J. and Shrestha, M. and Smith, N. and Dong, Y. and Meza Retamal, N. and Valenti, S. and Boos, S. and Shen, K. J. and Townsley, D. and Galbany, L. and Piscarreta, L. and Foley, R. J. and Bustamante-Rosell, M. J. and Coulter, D. A. and Chornock, R. and Davis, K. W. and Dickinson, C. B. and Jones, D. O. and Kutcka, J. and Le Saux, X. K. and Rojas-Bravo, C. R. and Taggart, K. and Tinyanont, S. and Yang, G. and Jha, S. W. and Margutti, R.},
title = {SN 2022joj: A Potential Double Detonation with a Thin Helium Shell},
journal = {The Astrophysical Journal}
}

@article{perley2019fast,
  title={The fast, luminous ultraviolet transient AT2018cow: extreme supernova, or disruption of a star by an intermediate-mass black hole?},
  author={Perley, Daniel A and Mazzali, Paolo A and Yan, Lin and Cenko, S Bradley and Gezari, Suvi and Taggart, Kirsty and Blagorodnova, Nadia and Fremling, Christoffer and Mockler, Brenna and Singh, Avinash and others},
  journal={Monthly Notices of the Royal Astronomical Society},
  volume={484},
  number={1},
  pages={1031--1049},
  year={2019},
  publisher={Oxford University Press}
}

@article{gal2019most,
  title={The most luminous supernovae},
  author={Gal-Yam, Avishay},
  journal={Annual Review of Astronomy and Astrophysics},
  volume={57},
  number={1},
  pages={305--333},
  year={2019},
  publisher={Annual Reviews}
}

@article{kasen2011pair,
  title={Pair instability supernovae: light curves, spectra, and shock breakout},
  author={Kasen, Daniel and Woosley, SE and Heger, Alexander},
  journal={The Astrophysical Journal},
  volume={734},
  number={2},
  pages={102},
  year={2011},
  publisher={The American Astronomical Society}
}

@article{karpenka2013simple,
  title={A simple and robust method for automated photometric classification of supernovae using neural networks},
  author={Karpenka, Natalia V and Feroz, F and Hobson, MP},
  journal={Monthly Notices of the Royal Astronomical Society},
  volume={429},
  number={2},
  pages={1278--1285},
  year={2013},
  publisher={The Royal Astronomical Society}
}

@article{phillips1993absolute,
  title={The absolute magnitudes of Type IA supernovae},
  author={Phillips, Mark M},
  journal={Astrophysical Journal, Part 2-Letters (ISSN 0004-637X), vol. 413, no. 2, p. L105-L108.},
  volume={413},
  pages={L105--L108},
  year={1993}
}

@article{law2022low,
  title={Low-cost access to the deep, high-cadence sky: the argus optical array},
  author={Law, Nicholas M and Corbett, Hank and Galliher, Nathan W and Gonzalez, Ramses and Vasquez, Alan and Walters, Glenn and Machia, Lawrence and Ratzloff, Jeff and Ackley, Kendall and Bizon, Chris and others},
  journal={Publications of the Astronomical Society of the Pacific},
  volume={134},
  number={1033},
  pages={035003},
  year={2022},
  publisher={The Astronomical Society of the Pacific}
}

@article{xu2025applecider,
  title={AppleCiDEr II: SpectraNet--A Deep Learning Network for Spectroscopic Data},
  author={Xu, Maojie and Sasli, Argyro and Junell, Alexandra and Nunes, Felipe Fontinele and Qin, Yu-Jing and Fremling, Christoffer and Rose, Sam and Laz, Theophile Jegou Du and Border, Benny and Calloch, Antoine Le and others},
  journal={arXiv preprint arXiv:2510.07215},
  year={2025}
}

@article{junell2026applying,
  title={Applying Multimodal Learning to Classify Transient Detections Early (AppleCiDEr). I. Dataset, Methods, and Infrastructure},
  author={Junell, Alexandra and Sasli, Argyro and Nunes, Felipe Fontinele and Xu, Maojie and Border, Benny and Rehemtulla, Nabeel and Rizhko, Mariia and Qin, Yu-Jing and Jegou Du Laz, Theophile and Le Calloch, Antoine and others},
  journal={Publications of the Astronomical Society of the Pacific},
  volume={138},
  number={5},
  pages={054508},
  year={2026},
  publisher={The Astronomical Society of the Pacific}
}

@inproceedings{devlin2019bert,
  title={Bert: Pre-training of deep bidirectional transformers for language understanding},
  author={Devlin, Jacob and Chang, Ming-Wei and Lee, Kenton and Toutanova, Kristina},
  booktitle={Proceedings of the 2019 conference of the North American chapter of the association for computational linguistics: human language technologies, volume 1 (long and short papers)},
  pages={4171--4186},
  year={2019}
}

@misc{anthropic2025claude,
  author       = {{Anthropic}},
  title        = {Claude Sonnet 4.6: Model Architecture and Training Methodology},
  year         = {2025},
  howpublished = {Technical report, Anthropic Transparency Hub},
  url          = {https://www.anthropic.com/transparency-hub}
}

@article{bi2024deepseek,
  title={Deepseek llm: Scaling open-source language models with longtermism},
  author={Bi, Xiao and Chen, Deli and Chen, Guanting and Chen, Shanhuang and Dai, Damai and Deng, Chengqi and Ding, Honghui and Dong, Kai and Du, Qiushi and Fu, Zhe and others},
  journal={arXiv preprint arXiv:2401.02954},
  year={2024}
}

@article{da2019tqdm,
  title={tqdm: A fast, extensible progress meter for python and cli},
  author={da Costa-Luis, Casper O},
  journal={Journal of Open Source Software},
  volume={4},
  number={37},
  pages={1277},
  year={2019},
  publisher={The Open Journal}
}

@article{scikit-learn,
     title={Scikit-learn: Machine Learning in {P}ython},
     author={Pedregosa, F. and Varoquaux, G. and Gramfort, A. and Michel, V.
             and Thirion, B. and Grisel, O. and Blondel, M. and Prettenhofer, P.
             and Weiss, R. and Dubourg, V. and Vanderplas, J. and Passos, A. and
             Cournapeau, D. and Brucher, M. and Perrot, M. and Duchesnay, E.},
     journal={Journal of Machine Learning Research},
     volume={12},
     pages={2825--2830},
     year={2011}
    }

@misc{kerashub2024,
  title={KerasHub},
  author={Watson, Matthew and  Chollet Fran\c{c}ois and Sreepathihalli,
  Divyashree and Saadat Samaneh and Sampath Ramesh and Rasskin Gabriel and
  and Zhu Scott and Singh Varun and Wood Luke and Tan Zhenyu and Stenbit,
  Ian and Qian Chen and Bischof Jonathan and others},
  year={2024},
  howpublished={\url{https://github.com/keras-team/keras-hub}},
}

@misc{chollet2015keras,
  title={Keras},
  author={Chollet, Fran\c{c}ois and others},
  year={2015},
  howpublished={\url{https://keras.io}},
}

@misc{tensorflow2015-whitepaper,
title={ {TensorFlow}: Large-Scale Machine Learning on Heterogeneous Systems},
url={https://www.tensorflow.org/},
note={Software available from tensorflow.org},
author={
    Mart\'{i}n~Abadi and
    Ashish~Agarwal and
    Paul~Barham and
    Eugene~Brevdo and
    Zhifeng~Chen and
    Craig~Citro and
    Greg~S.~Corrado and
    Andy~Davis and
    Jeffrey~Dean and
    Matthieu~Devin and
    Sanjay~Ghemawat and
    Ian~Goodfellow and
    Andrew~Harp and
    Geoffrey~Irving and
    Michael~Isard and
    Yangqing Jia and
    Rafal~Jozefowicz and
    Lukasz~Kaiser and
    Manjunath~Kudlur and
    Josh~Levenberg and
    Dandelion~Man\'{e} and
    Rajat~Monga and
    Sherry~Moore and
    Derek~Murray and
    Chris~Olah and
    Mike~Schuster and
    Jonathon~Shlens and
    Benoit~Steiner and
    Ilya~Sutskever and
    Kunal~Talwar and
    Paul~Tucker and
    Vincent~Vanhoucke and
    Vijay~Vasudevan and
    Fernanda~Vi\'{e}gas and
    Oriol~Vinyals and
    Pete~Warden and
    Martin~Wattenberg and
    Martin~Wicke and
    Yuan~Yu and
    Xiaoqiang~Zheng},
  year={2015},
}

@article{robitaille2013astropy,
  title={Astropy: A community Python package for astronomy},
  author={Robitaille, Thomas P and Tollerud, Erik J and Greenfield, Perry and Droettboom, Michael and Bray, Erik and Aldcroft, Tom and Davis, Matt and Ginsburg, Adam and Price-Whelan, Adrian M and Kerzendorf, Wolfgang E and others},
  journal={Astronomy \& Astrophysics},
  volume={558},
  pages={A33},
  year={2013},
  publisher={EDP Sciences}
}

@article{virtanen2020scipy,
  title={SciPy 1.0: fundamental algorithms for scientific computing in Python},
  author={Virtanen, Pauli and Gommers, Ralf and Oliphant, Travis E and Haberland, Matt and Reddy, Tyler and Cournapeau, David and Burovski, Evgeni and Peterson, Pearu and Weckesser, Warren and Bright, Jonathan and others},
  journal={Nature methods},
  volume={17},
  number={3},
  pages={261--272},
  year={2020},
  publisher={Nature Publishing Group US New York}
}

@ARTICLE{4160265,
  author={Hunter, John D.},
  journal={Computing in Science \& Engineering}, 
  title={Matplotlib: A 2D Graphics Environment}, 
  year={2007},
  volume={9},
  number={3},
  pages={90-95},
  doi={10.1109/MCSE.2007.55}}

@article{mckinney2010data,
  title={Data structures for statistical computing in Python.},
  author={McKinney, Wes and others},
  journal={scipy},
  volume={445},
  number={1},
  pages={51--56},
  year={2010}
}

@article{harris2020array,
  title={Array programming with NumPy},
  author={Harris, Charles R and Millman, K Jarrod and Van Der Walt, St{\'e}fan J and Gommers, Ralf and Virtanen, Pauli and Cournapeau, David and Wieser, Eric and Taylor, Julian and Berg, Sebastian and Smith, Nathaniel J and others},
  journal={nature},
  volume={585},
  number={7825},
  pages={357--362},
  year={2020},
  publisher={Nature Publishing Group UK London}
}

@INPROCEEDINGS{1997AAS...191.8504P,
       author = {{Perlmutter}, S. and {Aldering}, G. and {Deustua}, S. and {Fabbro}, S. and {Goldhaber}, G. and {Groom}, D.~E. and {Kim}, A.~G. and {Kim}, M.~Y. and {Knop}, R.~A. and {Nugent}, P. and {Pennypacker}, C.~R. and {della Valle}, M. and {Ellis}, R.~S. and {McMahon}, R.~G. and {Walton}, N. and {Fruchter}, A. and {Panagia}, N. and {Goobar}, A. and {Hook}, I.~M. and {Lidman}, C. and {Pain}, R. and {Ruiz-Lapuente}, P. and {Schaefer}, B. and {Supernova Cosmology Project}},
        title = "{Cosmology From Type IA Supernovae: Measurements, Calibration Techniques, and Implications}",
    booktitle = {American Astronomical Society Meeting Abstracts},
         year = 1997,
       series = {American Astronomical Society Meeting Abstracts},
       volume = {191},
        month = dec,
          eid = {85.04},
        pages = {85.04},
          doi = {10.48550/arXiv.astro-ph/9812473},
archivePrefix = {arXiv},
       eprint = {astro-ph/9812473},
 primaryClass = {astro-ph},
       adsurl = {https://ui.adsabs.harvard.edu/abs/1997AAS...191.8504P}
}

@ARTICLE{2019NatAs...3..717M,
       author = {{Modjaz}, Maryam and {Guti{\'e}rrez}, Claudia P. and {Arcavi}, Iair},
        title = "{New regimes in the observation of core-collapse supernovae}",
      journal = {Nature Astronomy},
         year = 2019,
        month = aug,
       volume = {3},
        pages = {717-724},
          doi = {10.1038/s41550-019-0856-2},
archivePrefix = {arXiv},
       eprint = {1908.02476},
 primaryClass = {astro-ph.HE},
       adsurl = {https://ui.adsabs.harvard.edu/abs/2019NatAs...3..717M}
}

@ARTICLE{1984SvA....28..658P,
       author = {{Pskovskii}, Yu. P.},
        title = "{Photometric classification and basic parameters of type I supernovae}",
      journal = {\sovast},
         year = 1984,
        month = dec,
       volume = {28},
        pages = {658-664},
       adsurl = {https://ui.adsabs.harvard.edu/abs/1984SvA....28..658P}
}

@INPROCEEDINGS{2012SPIE.8446E..86B,
       author = {{Ben-Ami}, Sagi and {Konidaris}, Nick and {Quimby}, Robert and {Davis}, Jack T. and {Ngeow}, Chow Choong and {Ritter}, Andreas and {Rudy}, Alexander},
        title = "{The SED Machine: a dedicated transient IFU spectrograph}",
    booktitle = {Ground-based and Airborne Instrumentation for Astronomy IV},
         year = 2012,
       editor = {{McLean}, Ian S. and {Ramsay}, Suzanne K. and {Takami}, Hideki},
       series = {Society of Photo-Optical Instrumentation Engineers (SPIE) Conference Series},
       volume = {8446},
        month = sep,
          eid = {844686},
        pages = {844686},
          doi = {10.1117/12.926317},
       adsurl = {https://ui.adsabs.harvard.edu/abs/2012SPIE.8446E..86B}
}

@article{Stoppa2026,
    author = {Stoppa, Fiorenzo and Smartt, Stephen J},
    title = {SNID–SAGE: a modern framework for interactive supernova classification and spectral analysis},
    journal = {Monthly Notices of the Royal Astronomical Society},
    volume = {549},
    number = {4},
    pages = {stag1066},
    year = {2026},
    month = {07},
    issn = {0035-8711},
    doi = {10.1093/mnras/stag1066},
    url = {https://doi.org/10.1093/mnras/stag1066},
    eprint = {https://academic.oup.com/mnras/article-pdf/549/4/stag1066/68483078/stag1066.pdf},
}

@article{shivvers2017revisiting,
  title={Revisiting the lick observatory supernova search volume-limited sample: updated classifications and revised stripped-envelope supernova fractions},
  author={Shivvers, Isaac and Modjaz, Maryam and Zheng, WeiKang and Liu, Yuqian and Filippenko, Alexei V and Silverman, Jeffrey M and Matheson, Thomas and Pastorello, Andrea and Graur, Or and Foley, Ryan J and others},
  journal={Publications of the Astronomical Society of the Pacific},
  volume={129},
  number={975},
  pages={054201},
  year={2017},
  publisher={The Astronomical Society of the Pacific}
}

@article{patat2001metamorphosis,
  title={The metamorphosis of SN 1998bw},
  author={Patat, Ferdinando and Cappellaro, Enrico and Danziger, John and Mazzali, Paolo A and Sollerman, Jesper and Augusteijn, Thomas and Brewer, James and Doublier, Vanessa and Gonzalez, Jean Francois and Hainaut, Olivier and others},
  journal={The Astrophysical Journal},
  volume={555},
  number={2},
  pages={900--917},
  year={2001}
}

@article{dessart2012nature,
  title={On the nature of supernovae Ib and Ic},
  author={Dessart, Luc and Hillier, D John and Li, Chengdong and Woosley, Stan},
  journal={Monthly Notices of the Royal Astronomical Society},
  volume={424},
  number={3},
  pages={2139--2159},
  year={2012},
  publisher={Blackwell Science Ltd Oxford, UK}
}

@article{sauer2006properties,
  title={The properties of the ‘standard’Type Ic supernova 1994I from spectral models},
  author={Sauer, Daniel N and Mazzali, PA and Deng, J and Valenti, S and Nomoto, K and Filippenko, AV},
  journal={Monthly Notices of the Royal Astronomical Society},
  volume={369},
  number={4},
  pages={1939--1948},
  year={2006},
  publisher={Blackwell Publishing Ltd Oxford, UK}
}

@ARTICLE{1996ApJ...462..462C,
       author = {{Clocchiatti}, Alejandro and {Wheeler}, J. Craig and {Brotherton}, Michael S. and {Cochran}, Anita L. and {Wills}, Derek and {Barker}, Edwin S. and {Turatto}, Massimo},
        title = "{SN 1994I: Disentangling He i Lines in Type IC Supernovae}",
      journal = {\apj},
         year = 1996,
        month = may,
       volume = {462},
        pages = {462},
          doi = {10.1086/177165},
       adsurl = {https://ui.adsabs.harvard.edu/abs/1996ApJ...462..462C}
}

@article{liu_type_2023,
	title = {Type {Ia} {Supernova} {Explosions} in {Binary} {Systems}: {A} {Review}},
	volume = {23},
	issn = {1674-4527},
	shorttitle = {Type {Ia} {Supernova} {Explosions} in {Binary} {Systems}},
	url = {https://doi.org/10.1088/1674-4527/acd89e},
	doi = {10.1088/1674-4527/acd89e},
	language = {en},
	number = {8},
	urldate = {2026-07-01},
	journal = {Research in Astronomy and Astrophysics},
	publisher = {National Astromonical Observatories, CAS and IOP Publishing},
	author = {Liu, Zheng-Wei and Röpke, Friedrich K. and Han, Zhanwen},
	month = jul,
	year = {2023},
	pages = {082001},
}

@article{williamson_modeling_2021,
	title = {Modeling {Type} {Ic} {Supernovae} with {TARDIS}: {Hidden} {Helium} in {SN} {1994I}?},
	volume = {908},
	issn = {0004-637X},
	shorttitle = {Modeling {Type} {Ic} {Supernovae} with {TARDIS}},
	url = {https://ui.adsabs.harvard.edu/abs/2021ApJ...908..150W},
	doi = {10.3847/1538-4357/abd244},
	urldate = {2026-07-01},
	journal = {The Astrophysical Journal},
	publisher = {IOP},
	author = {Williamson, Marc and Kerzendorf, Wolfgang and Modjaz, Maryam},
	month = feb,
	year = {2021},
	note = {ADS Bibcode: 2021ApJ...908..150W},
	pages = {150},
}

@article{chen_sn_2018,
	title = {{SN} 2017ens: {The} {Metamorphosis} of a {Luminous} {Broadlined} {Type} {Ic} {Supernova} into an {SN} {IIn}},
	volume = {867},
	issn = {2041-8205},
	shorttitle = {{SN} 2017ens},
	url = {https://doi.org/10.3847/2041-8213/aaeb2e},
	doi = {10.3847/2041-8213/aaeb2e},
	language = {en},
	number = {2},
	urldate = {2026-07-01},
	journal = {The Astrophysical Journal Letters},
	publisher = {The American Astronomical Society},
	author = {Chen, T.-W. and Inserra, C. and Fraser, M. and Moriya, T. J. and Schady, P. and Schweyer, T. and Filippenko, A. V. and Perley, D. A. and Ruiter, A. J. and Seitenzahl, I. and Sollerman, J. and Taddia, F. and Anderson, J. P. and Foley, R. J. and Jerkstrand, A. and Ngeow, C.-C. and Pan, Y.-C. and Pastorello, A. and Points, S. and Smartt, S. J. and Smith, K. W. and Taubenberger, S. and Wiseman, P. and Young, D. R. and Benetti, S. and Berton, M. and Bufano, F. and Clark, P. and Valle, M. Della and Galbany, L. and Gal-Yam, A. and Gromadzki, M. and Gutiérrez, C. P. and Heinze, A. and Kankare, E. and Kilpatrick, C. D. and Kuncarayakti, H. and Leloudas, G. and Lin, Z.-Y. and Maguire, K. and Mazzali, P. and McBrien, O. and Prentice, S. J. and Rau, A. and Rest, A. and Siebert, M. R. and Stalder, B. and Tonry, J. L. and Yu, P.-C.},
	month = nov,
	year = {2018},
	pages = {L31},
}

@article{chandra_multiwavelength_2025,
	title = {Multiwavelength {View} of {Circumstellar} {Interaction} in {Supernovae}},
	volume = {11},
	url = {https://ui.adsabs.harvard.edu/abs/2025Univ...11..363C},
	doi = {10.3390/universe11110363},
	urldate = {2026-07-01},
	journal = {Universe},
	publisher = {MDPI},
	author = {Chandra, Poonam},
	month = nov,
	year = {2025},
	note = {ADS Bibcode: 2025Univ...11..363C},
	pages = {363},
}

@article{woosley_supernova_2006,
	title = {The {Supernova} {Gamma}-{Ray} {Burst} {Connection}},
	volume = {44},
	issn = {0066-4146},
	url = {https://ui.adsabs.harvard.edu/abs/2006ARA&A..44..507W},
	doi = {10.1146/annurev.astro.43.072103.150558},
	urldate = {2026-07-01},
	journal = {Annual Review of Astronomy and Astrophysics},
	author = {Woosley, S. E. and Bloom, J. S.},
	month = sep,
	year = {2006},
	note = {ADS Bibcode: 2006ARA\&A..44..507W},
	pages = {507--556},
}

@article{iwamoto_peculiar_2000,
	title = {The {Peculiar} {Type} {Ic} {Supernova} 1997ef:{Another} {Hypernova}},
	volume = {534},
	issn = {0004-637X},
	shorttitle = {The {Peculiar} {Type} {Ic} {Supernova} 1997ef},
	url = {https://iopscience.iop.org/article/10.1086/308761},
	doi = {10.1086/308761},
	language = {en},
	number = {2},
	urldate = {2026-07-01},
	journal = {The Astrophysical Journal},
	publisher = {IOP Publishing},
	author = {Iwamoto, Koichi and Nakamura, Takayoshi and Nomoto, Ken’ichi and Mazzali, Paolo A. and Danziger, I. John and Garnavich, Peter and Kirshner, Robert and Jha, Saurabh and Balam, David and Thorstensen, John},
	month = may,
	year = {2000},
	pages = {660},
}

@article{foley_type_2013,
	title = {{TYPE} {Iax} {SUPERNOVAE}: {A} {NEW} {CLASS} {OF} {STELLAR} {EXPLOSION}*},
	volume = {767},
	issn = {0004-637X},
	shorttitle = {{TYPE} {Iax} {SUPERNOVAE}},
	url = {https://doi.org/10.1088/0004-637X/767/1/57},
	doi = {10.1088/0004-637X/767/1/57},
	language = {en},
	number = {1},
	urldate = {2026-06-04},
	journal = {The Astrophysical Journal},
	publisher = {The American Astronomical Society},
	author = {Foley, Ryan J. and Challis, P. J. and Chornock, R. and Ganeshalingam, M. and Li, W. and Marion, G. H. and Morrell, N. I. and Pignata, G. and Stritzinger, M. D. and Silverman, J. M. and Wang, X. and Anderson, J. P. and Filippenko, A. V. and Freedman, W. L. and Hamuy, M. and Jha, S. W. and Kirshner, R. P. and McCully, C. and Persson, S. E. and Phillips, M. M. and Reichart, D. E. and Soderberg, A. M.},
	month = mar,
	year = {2013},
	pages = {57},
}

@article{ivezic_lsst_2019,
	title = {{LSST}: {From} {Science} {Drivers} to {Reference} {Design} and {Anticipated} {Data} {Products}},
	volume = {873},
	issn = {0004-637X},
	shorttitle = {{LSST}},
	url = {https://ui.adsabs.harvard.edu/abs/2019ApJ...873..111I},
	doi = {10.3847/1538-4357/ab042c},
	urldate = {2022-11-08},
	journal = {The Astrophysical Journal},
	author = {Ivezić, Željko and Kahn, Steven M. and Tyson, J. Anthony and Abel, Bob and Acosta, Emily and Allsman, Robyn and Alonso, David and AlSayyad, Yusra and Anderson, Scott F. and Andrew, John and Angel, James Roger P. and Angeli, George Z. and Ansari, Reza and Antilogus, Pierre and Araujo, Constanza and Armstrong, Robert and Arndt, Kirk T. and Astier, Pierre and Aubourg, Éric and Auza, Nicole and Axelrod, Tim S. and Bard, Deborah J. and Barr, Jeff D. and Barrau, Aurelian and Bartlett, James G. and Bauer, Amanda E. and Bauman, Brian J. and Baumont, Sylvain and Bechtol, Ellen and Bechtol, Keith and Becker, Andrew C. and Becla, Jacek and Beldica, Cristina and Bellavia, Steve and Bianco, Federica B. and Biswas, Rahul and Blanc, Guillaume and Blazek, Jonathan and Blandford, Roger D. and Bloom, Josh S. and Bogart, Joanne and Bond, Tim W. and Booth, Michael T. and Borgland, Anders W. and Borne, Kirk and Bosch, James F. and Boutigny, Dominique and Brackett, Craig A. and Bradshaw, Andrew and Brandt, William Nielsen and Brown, Michael E. and Bullock, James S. and Burchat, Patricia and Burke, David L. and Cagnoli, Gianpietro and Calabrese, Daniel and Callahan, Shawn and Callen, Alice L. and Carlin, Jeffrey L. and Carlson, Erin L. and Chandrasekharan, Srinivasan and Charles-Emerson, Glenaver and Chesley, Steve and Cheu, Elliott C. and Chiang, Hsin-Fang and Chiang, James and Chirino, Carol and Chow, Derek and Ciardi, David R. and Claver, Charles F. and Cohen-Tanugi, Johann and Cockrum, Joseph J. and Coles, Rebecca and Connolly, Andrew J. and Cook, Kem H. and Cooray, Asantha and Covey, Kevin R. and Cribbs, Chris and Cui, Wei and Cutri, Roc and Daly, Philip N. and Daniel, Scott F. and Daruich, Felipe and Daubard, Guillaume and Daues, Greg and Dawson, William and Delgado, Francisco and Dellapenna, Alfred and de Peyster, Robert and de Val-Borro, Miguel and Digel, Seth W. and Doherty, Peter and Dubois, Richard and Dubois-Felsmann, Gregory P. and Durech, Josef and Economou, Frossie and Eifler, Tim and Eracleous, Michael and Emmons, Benjamin L. and Fausti Neto, Angelo and Ferguson, Henry and Figueroa, Enrique and Fisher-Levine, Merlin and Focke, Warren and Foss, Michael D. and Frank, James and Freemon, Michael D. and Gangler, Emmanuel and Gawiser, Eric and Geary, John C. and Gee, Perry and Geha, Marla and Gessner, Charles J. B. and Gibson, Robert R. and Gilmore, D. Kirk and Glanzman, Thomas and Glick, William and Goldina, Tatiana and Goldstein, Daniel A. and Goodenow, Iain and Graham, Melissa L. and Gressler, William J. and Gris, Philippe and Guy, Leanne P. and Guyonnet, Augustin and Haller, Gunther and Harris, Ron and Hascall, Patrick A. and Haupt, Justine and Hernandez, Fabio and Herrmann, Sven and Hileman, Edward and Hoblitt, Joshua and Hodgson, John A. and Hogan, Craig and Howard, James D. and Huang, Dajun and Huffer, Michael E. and Ingraham, Patrick and Innes, Walter R. and Jacoby, Suzanne H. and Jain, Bhuvnesh and Jammes, Fabrice and Jee, M. James and Jenness, Tim and Jernigan, Garrett and Jevremović, Darko and Johns, Kenneth and Johnson, Anthony S. and Johnson, Margaret W. G. and Jones, R. Lynne and Juramy-Gilles, Claire and Jurić, Mario and Kalirai, Jason S. and Kallivayalil, Nitya J. and Kalmbach, Bryce and Kantor, Jeffrey P. and Karst, Pierre and Kasliwal, Mansi M. and Kelly, Heather and Kessler, Richard and Kinnison, Veronica and Kirkby, David and Knox, Lloyd and Kotov, Ivan V. and Krabbendam, Victor L. and Krughoff, K. Simon and Kubánek, Petr and Kuczewski, John and Kulkarni, Shri and Ku, John and Kurita, Nadine R. and Lage, Craig S. and Lambert, Ron and Lange, Travis and Langton, J. Brian and Le Guillou, Laurent and Levine, Deborah and Liang, Ming and Lim, Kian-Tat and Lintott, Chris J. and Long, Kevin E. and Lopez, Margaux and Lotz, Paul J. and Lupton, Robert H. and Lust, Nate B. and MacArthur, Lauren A. and Mahabal, Ashish and Mandelbaum, Rachel and Markiewicz, Thomas W. and Marsh, Darren S. and Marshall, Philip J. and Marshall, Stuart and May, Morgan and McKercher, Robert and McQueen, Michelle and Meyers, Joshua and Migliore, Myriam and Miller, Michelle and Mills, David J. and Miraval, Connor and Moeyens, Joachim and Moolekamp, Fred E. and Monet, David G. and Moniez, Marc and Monkewitz, Serge and Montgomery, Christopher and Morrison, Christopher B. and Mueller, Fritz and Muller, Gary P. and Muñoz Arancibia, Freddy and Neill, Douglas R. and Newbry, Scott P. and Nief, Jean-Yves and Nomerotski, Andrei and Nordby, Martin and O'Connor, Paul and Oliver, John and Olivier, Scot S. and Olsen, Knut and O'Mullane, William and Ortiz, Sandra and Osier, Shawn and Owen, Russell E. and Pain, Reynald and Palecek, Paul E. and Parejko, John K. and Parsons, James B. and Pease, Nathan M. and Peterson, J. Matt and Peterson, John R. and Petravick, Donald L. and Libby Petrick, M. E. and Petry, Cathy E. and Pierfederici, Francesco and Pietrowicz, Stephen and Pike, Rob and Pinto, Philip A. and Plante, Raymond and Plate, Stephen and Plutchak, Joel P. and Price, Paul A. and Prouza, Michael and Radeka, Veljko and Rajagopal, Jayadev and Rasmussen, Andrew P. and Regnault, Nicolas and Reil, Kevin A. and Reiss, David J. and Reuter, Michael A. and Ridgway, Stephen T. and Riot, Vincent J. and Ritz, Steve and Robinson, Sean and Roby, William and Roodman, Aaron and Rosing, Wayne and Roucelle, Cecille and Rumore, Matthew R. and Russo, Stefano and Saha, Abhijit and Sassolas, Benoit and Schalk, Terry L. and Schellart, Pim and Schindler, Rafe H. and Schmidt, Samuel and Schneider, Donald P. and Schneider, Michael D. and Schoening, William and Schumacher, German and Schwamb, Megan E. and Sebag, Jacques and Selvy, Brian and Sembroski, Glenn H. and Seppala, Lynn G. and Serio, Andrew and Serrano, Eduardo and Shaw, Richard A. and Shipsey, Ian and Sick, Jonathan and Silvestri, Nicole and Slater, Colin T. and Smith, J. Allyn and Smith, R. Chris and Sobhani, Shahram and Soldahl, Christine and Storrie-Lombardi, Lisa and Stover, Edward and Strauss, Michael A. and Street, Rachel A. and Stubbs, Christopher W. and Sullivan, Ian S. and Sweeney, Donald and Swinbank, John D. and Szalay, Alexander and Takacs, Peter and Tether, Stephen A. and Thaler, Jon J. and Thayer, John Gregg and Thomas, Sandrine and Thornton, Adam J. and Thukral, Vaikunth and Tice, Jeffrey and Trilling, David E. and Turri, Max and Van Berg, Richard and Vanden Berk, Daniel and Vetter, Kurt and Virieux, Francoise and Vucina, Tomislav and Wahl, William and Walkowicz, Lucianne and Walsh, Brian and Walter, Christopher W. and Wang, Daniel L. and Wang, Shin-Yawn and Warner, Michael and Wiecha, Oliver and Willman, Beth and Winters, Scott E. and Wittman, David and Wolff, Sidney C. and Wood-Vasey, W. Michael and Wu, Xiuqin and Xin, Bo and Yoachim, Peter and Zhan, Hu},
	month = mar,
	year = {2019},
	note = {ADS Bibcode: 2019ApJ...873..111I},
	pages = {111},
}

@article{gal-yam_most_2019,
	title = {The {Most} {Luminous} {Supernovae}},
	volume = {57},
	issn = {0066-4146, 1545-4282},
	url = {https://www.annualreviews.org/content/journals/10.1146/annurev-astro-081817-051819},
	doi = {10.1146/annurev-astro-081817-051819},
	language = {en},
	number = {Volume 57, 2019},
	urldate = {2026-06-17},
	journal = {Annual Review of Astronomy and Astrophysics},
	publisher = {Annual Reviews},
	author = {Gal-Yam, Avishay},
	month = aug,
	year = {2019},
	pages = {305--333},
}

@article{folatelli_spectral_2004,
	series = {Proceedings of the {Workshop} on {Supernovae} and {Dust}},
	title = {Spectral homogeneity of type {Ia} supernovae},
	volume = {48},
	issn = {1387-6473},
	url = {https://www.sciencedirect.com/science/article/pii/S1387647303004056},
	doi = {10.1016/j.newar.2003.12.039},
	number = {7},
	urldate = {2026-06-12},
	journal = {New Astronomy Reviews},
	author = {Folatelli, G.},
	month = may,
	year = {2004},
	pages = {623--628},
}

@article{fryer_compact_2013,
	title = {Compact object formation and the supernova explosion engine},
	volume = {30},
	issn = {0264-9381},
	url = {https://doi.org/10.1088/0264-9381/30/24/244002},
	doi = {10.1088/0264-9381/30/24/244002},
	language = {en},
	number = {24},
	urldate = {2026-06-10},
	journal = {Classical and Quantum Gravity},
	publisher = {IOP Publishing},
	author = {Fryer, Chris L},
	month = nov,
	year = {2013},
	pages = {244002},
}

@article{qu_scone_2021,
	title = {{SCONE}: {Supernova} {Classification} with a {Convolutional} {Neural} {Network}},
	volume = {162},
	issn = {0004-6256},
	shorttitle = {{SCONE}},
	url = {https://ui.adsabs.harvard.edu/abs/2021AJ....162...67Q},
	doi = {10.3847/1538-3881/ac0824},
	urldate = {2026-06-09},
	journal = {The Astronomical Journal},
	publisher = {IOP},
	author = {Qu, Helen and Sako, Masao and Möller, Anais and Doux, Cyrille},
	month = aug,
	year = {2021},
	note = {ADS Bibcode: 2021AJ....162...67Q},
	pages = {67},
}

@article{moller_supernnova_2020,
	title = {{SuperNNova}: an open-source framework for {Bayesian}, neural network-based supernova classification},
	volume = {491},
	issn = {0035-8711},
	shorttitle = {{SuperNNova}},
	url = {https://ui.adsabs.harvard.edu/abs/2020MNRAS.491.4277M},
	doi = {10.1093/mnras/stz3312},
	urldate = {2026-06-09},
	journal = {Monthly Notices of the Royal Astronomical Society},
	publisher = {OUP},
	author = {Möller, A. and de Boissière, T.},
	month = jan,
	year = {2020},
	note = {ADS Bibcode: 2020MNRAS.491.4277M},
	pages = {4277--4293},
}

@article{boone_avocado_2019,
	title = {Avocado: {Photometric} {Classification} of {Astronomical} {Transients} with {Gaussian} {Process} {Augmentation}},
	volume = {158},
	issn = {0004-6256},
	shorttitle = {Avocado},
	url = {https://ui.adsabs.harvard.edu/abs/2019AJ....158..257B},
	doi = {10.3847/1538-3881/ab5182},
	urldate = {2026-06-09},
	journal = {The Astronomical Journal},
	publisher = {IOP},
	author = {Boone, Kyle},
	month = dec,
	year = {2019},
	note = {ADS Bibcode: 2019AJ....158..257B},
	pages = {257},
}

@article{bellm_zwicky_2018,
	title = {The {Zwicky} {Transient} {Facility}: {System} {Overview}, {Performance}, and {First} {Results}},
	volume = {131},
	issn = {1538-3873},
	shorttitle = {The {Zwicky} {Transient} {Facility}},
	url = {https://doi.org/10.1088/1538-3873/aaecbe},
	doi = {10.1088/1538-3873/aaecbe},
	language = {en},
	number = {995},
	urldate = {2026-06-08},
	journal = {Publications of the Astronomical Society of the Pacific},
	publisher = {The Astronomical Society of the Pacific},
	author = {Bellm, Eric C. and Kulkarni, Shrinivas R. and Graham, Matthew J. and Dekany, Richard and Smith, Roger M. and Riddle, Reed and Masci, Frank J. and Helou, George and Prince, Thomas A. and Adams, Scott M. and Barbarino, C. and Barlow, Tom and Bauer, James and Beck, Ron and Belicki, Justin and Biswas, Rahul and Blagorodnova, Nadejda and Bodewits, Dennis and Bolin, Bryce and Brinnel, Valery and Brooke, Tim and Bue, Brian and Bulla, Mattia and Burruss, Rick and Cenko, S. Bradley and Chang, Chan-Kao and Connolly, Andrew and Coughlin, Michael and Cromer, John and Cunningham, Virginia and De, Kishalay and Delacroix, Alex and Desai, Vandana and Duev, Dmitry A. and Eadie, Gwendolyn and Farnham, Tony L. and Feeney, Michael and Feindt, Ulrich and Flynn, David and Franckowiak, Anna and Frederick, S. and Fremling, C. and Gal-Yam, Avishay and Gezari, Suvi and Giomi, Matteo and Goldstein, Daniel A. and Golkhou, V. Zach and Goobar, Ariel and Groom, Steven and Hacopians, Eugean and Hale, David and Henning, John and Ho, Anna Y. Q. and Hover, David and Howell, Justin and Hung, Tiara and Huppenkothen, Daniela and Imel, David and Ip, Wing-Huen and Ivezić, Željko and Jackson, Edward and Jones, Lynne and Juric, Mario and Kasliwal, Mansi M. and Kaspi, S. and Kaye, Stephen and Kelley, Michael S. P. and Kowalski, Marek and Kramer, Emily and Kupfer, Thomas and Landry, Walter and Laher, Russ R. and Lee, Chien-De and Lin, Hsing Wen and Lin, Zhong-Yi and Lunnan, Ragnhild and Giomi, Matteo and Mahabal, Ashish and Mao, Peter and Miller, Adam A. and Monkewitz, Serge and Murphy, Patrick and Ngeow, Chow-Choong and Nordin, Jakob and Nugent, Peter and Ofek, Eran and Patterson, Maria T. and Penprase, Bryan and Porter, Michael and Rauch, Ludwig and Rebbapragada, Umaa and Reiley, Dan and Rigault, Mickael and Rodriguez, Hector and Roestel, Jan van and Rusholme, Ben and Santen, Jakob van and Schulze, S. and Shupe, David L. and Singer, Leo P. and Soumagnac, Maayane T. and Stein, Robert and Surace, Jason and Sollerman, Jesper and Szkody, Paula and Taddia, F. and Terek, Scott and Van Sistine, Angela and van Velzen, Sjoert and Vestrand, W. Thomas and Walters, Richard and Ward, Charlotte and Ye, Quan-Zhi and Yu, Po-Chieh and Yan, Lin and Zolkower, Jeffry},
	month = dec,
	year = {2018},
	pages = {018002},
}

@article{colgate_early_1969,
	title = {Early {Supernova} {Luminosity}},
	volume = {157},
	issn = {0004-637X},
	url = {https://ui.adsabs.harvard.edu/abs/1969ApJ...157..623C},
	doi = {10.1086/150102},
	urldate = {2026-06-08},
	journal = {The Astrophysical Journal},
	publisher = {IOP},
	author = {Colgate, Stirling A. and McKee, Chester},
	month = aug,
	year = {1969},
	note = {ADS Bibcode: 1969ApJ...157..623C},
	pages = {623},
}

@article{diehl_sn2014j_2015,
	title = {{SN2014J} gamma rays from the {56Ni} decay chain},
	volume = {574},
	copyright = {© ESO, 2015},
	issn = {0004-6361, 1432-0746},
	url = {https://www.aanda.org/articles/aa/abs/2015/02/aa24991-14/aa24991-14.html},
	doi = {10.1051/0004-6361/201424991},
	language = {en},
	urldate = {2026-06-08},
	journal = {Astronomy \& Astrophysics},
	publisher = {EDP Sciences},
	author = {Diehl, Roland and Siegert, Thomas and Hillebrandt, Wolfgang and Krause, Martin and Greiner, Jochen and Maeda, Keiichi and Röpke, Friedrich K. and Sim, Stuart A. and Wang, Wei and Zhang, Xiaoling},
	month = feb,
	year = {2015},
	pages = {A72},
}

@article{chandrasekhar_maximum_1931,
	title = {The {Maximum} {Mass} of {Ideal} {White} {Dwarfs}},
	volume = {74},
	issn = {0004-637X},
	url = {https://ui.adsabs.harvard.edu/abs/1931ApJ....74...81C},
	doi = {10.1086/143324},
	urldate = {2026-06-08},
	journal = {The Astrophysical Journal},
	publisher = {IOP},
	author = {Chandrasekhar, S.},
	month = jul,
	year = {1931},
	note = {ADS Bibcode: 1931ApJ....74...81C},
	pages = {81},
}

@article{smartt_progenitors_2009,
	title = {Progenitors of {Core}-{Collapse} {Supernovae}},
	volume = {47},
	issn = {0066-4146},
	url = {https://ui.adsabs.harvard.edu/abs/2009ARA&A..47...63S},
	doi = {10.1146/annurev-astro-082708-101737},
	urldate = {2026-06-08},
	journal = {Annual Review of Astronomy and Astrophysics},
	author = {Smartt, Stephen J.},
	month = sep,
	year = {2009},
	note = {ADS Bibcode: 2009ARA\&A..47...63S},
	pages = {63--106},
}

@article{modjaz_early-time_2006,
	title = {Early-{Time} {Photometry} and {Spectroscopy} of the {Fast} {Evolving} {SN} 2006aj {Associated} with {GRB} 060218*},
	volume = {645},
	issn = {0004-637X},
	url = {https://iopscience.iop.org/article/10.1086/505906},
	doi = {10.1086/505906},
	language = {en},
	number = {1},
	urldate = {2026-06-06},
	journal = {The Astrophysical Journal},
	publisher = {IOP Publishing},
	author = {Modjaz, M. and Stanek, K. Z. and Garnavich, P. M. and Berlind, P. and Blondin, S. and Brown, W. and Calkins, M. and Challis, P. and Diamond-Stanic, A. M. and Hao, H. and Hicken, M. and Kirshner, R. P. and Prieto, J. L.},
	month = jun,
	year = {2006},
	pages = {L21},
}

@article{filippenko_supernova_1988,
	title = {Supernova {1987K}: {Type} {II} in {Youth}, {Type} {Ib} in {Old} {Age}},
	volume = {96},
	issn = {0004-6256},
	shorttitle = {Supernova {1987K}},
	url = {https://ui.adsabs.harvard.edu/abs/1988AJ.....96.1941F},
	doi = {10.1086/114940},
	urldate = {2026-06-04},
	journal = {The Astronomical Journal},
	publisher = {IOP},
	author = {Filippenko, Alexei V.},
	month = dec,
	year = {1988},
	note = {ADS Bibcode: 1988AJ.....96.1941F},
	pages = {1941},
}

@article{tripp_using_1997,
	title = {Using distant type {IA} supernovae to measure the cosmological expansion parameters.},
	volume = {325},
	issn = {0004-6361},
	url = {https://ui.adsabs.harvard.edu/abs/1997A&A...325..871T},
	urldate = {2026-06-03},
	journal = {Astronomy and Astrophysics},
	publisher = {EDP},
	author = {Tripp, R.},
	month = sep,
	year = {1997},
	note = {ADS Bibcode: 1997A\&A...325..871T},
	pages = {871--876},
}

@article{riess_precise_1996,
	title = {A {Precise} {Distance} {Indicator}: {Type} {IA} {Supernova} {Multicolor} {Light}-{Curve} {Shapes}},
	volume = {473},
	issn = {0004-637X},
	shorttitle = {A {Precise} {Distance} {Indicator}},
	url = {https://ui.adsabs.harvard.edu/abs/1996ApJ...473...88R},
	doi = {10.1086/178129},
	urldate = {2026-06-03},
	journal = {The Astrophysical Journal},
	publisher = {IOP},
	author = {Riess, Adam G. and Press, William H. and Kirshner, Robert P.},
	month = dec,
	year = {1996},
	note = {ADS Bibcode: 1996ApJ...473...88R},
	pages = {88},
}

@article{modjaz_shock_2009,
	title = {{FROM} {SHOCK} {BREAKOUT} {TO} {PEAK} {AND} {BEYOND}: {EXTENSIVE} {PANCHROMATIC} {OBSERVATIONS} {OF} {THE} {TYPE} {Ib} {SUPERNOVA} {2008D} {ASSOCIATED} {WITH} {SWIFT} {X}-{RAY} {TRANSIENT} 080109},
	volume = {702},
	issn = {0004-637X},
	shorttitle = {{FROM} {SHOCK} {BREAKOUT} {TO} {PEAK} {AND} {BEYOND}},
	url = {https://doi.org/10.1088/0004-637X/702/1/226},
	doi = {10.1088/0004-637X/702/1/226},
	language = {en},
	number = {1},
	urldate = {2026-06-03},
	journal = {The Astrophysical Journal},
	publisher = {The American Astronomical Society},
	author = {Modjaz, M. and Li, W. and Butler, N. and Chornock, R. and Perley, D. and Blondin, S. and Bloom, J. S. and Filippenko, A. V. and Kirshner, R. P. and Kocevski, D. and Poznanski, D. and Hicken, M. and Foley, R. J. and Stringfellow, G. S. and Berlind, P. and Barrado y Navascues, D. and Blake, C. H. and Bouy, H. and Brown, W. R. and Challis, P. and Chen, H. and de Vries, W. H. and Dufour, P. and Falco, E. and Friedman, A. and Ganeshalingam, M. and Garnavich, P. and Holden, B. and Illingworth, G. and Lee, N. and Liebert, J. and Marion, G. H. and Olivier, S. S. and Prochaska, J. X. and Silverman, J. M. and Smith, N. and Starr, D. and Steele, T. N. and Stockton, A. and Williams, G. G. and Wood-Vasey, W. M.},
	month = aug,
	year = {2009},
	pages = {226},
}

@article{wallerstein_synthesis_1997,
	title = {Synthesis of the elements in stars: forty years of progress},
	volume = {69},
	shorttitle = {Synthesis of the elements in stars},
	url = {https://link.aps.org/doi/10.1103/RevModPhys.69.995},
	doi = {10.1103/RevModPhys.69.995},
	number = {4},
	urldate = {2026-06-03},
	journal = {Reviews of Modern Physics},
	publisher = {American Physical Society},
	author = {Wallerstein, George and Iben, Icko and Parker, Peter and Boesgaard, Ann Merchant and Hale, Gerald M. and Champagne, Arthur E. and Barnes, Charles A. and Käppeler, Franz and Smith, Verne V. and Hoffman, Robert D. and Timmes, Frank X. and Sneden, Chris and Boyd, Richard N. and Meyer, Bradley S. and Lambert, David L.},
	month = oct,
	year = {1997},
	pages = {995--1084},
}

@article{burbidge_synthesis_1957,
	title = {Synthesis of the {Elements} in {Stars}},
	volume = {29},
	issn = {0034-6861},
	url = {https://ui.adsabs.harvard.edu/abs/1957RvMP...29..547B},
	doi = {10.1103/RevModPhys.29.547},
	urldate = {2026-06-02},
	journal = {Reviews of Modern Physics},
	publisher = {APS},
	author = {Burbidge, E. Margaret and Burbidge, G. R. and Fowler, William A. and Hoyle, F.},
	month = oct,
	year = {1957},
	note = {ADS Bibcode: 1957RvMP...29..547B},
	pages = {547--650},
}

@article{fortino_abc-sn_2026,
	title = {{ABC}-{SN}: {Attention}-based {Classifier} for {Supernova} {Spectra}},
	volume = {1000},
	issn = {0004-637X},
	shorttitle = {{ABC}-{SN}},
	url = {https://ui.adsabs.harvard.edu/abs/2026ApJ..1000...14F},
	doi = {10.3847/1538-4357/ae3b41},
	urldate = {2026-05-25},
	journal = {The Astrophysical Journal},
	publisher = {IOP},
	author = {Fortino, Willow Fox and Bianco, Federica B. and Protopapas, Pavlos and Muthukrishna, Daniel and Brockmeier, Austin},
	month = mar,
	year = {2026},
	note = {ADS Bibcode: 2026ApJ..1000...14F},
	pages = {14},
}

@article{li_sn_2003,
	title = {{SN} 2002cx: {The} {Most} {Peculiar} {Known} {Type} {Ia} {Supernova}},
	volume = {115},
	issn = {0004-6280},
	shorttitle = {{SN} 2002cx},
	url = {https://ui.adsabs.harvard.edu/abs/2003PASP..115..453L},
	doi = {10.1086/374200},
	urldate = {2026-03-11},
	journal = {Publications of the Astronomical Society of the Pacific},
	publisher = {IOP},
	author = {Li, Weidong and Filippenko, Alexei V. and Chornock, Ryan and Berger, Edo and Berlind, Perry and Calkins, Michael L. and Challis, Peter and Fassnacht, Chris and Jha, Saurabh and Kirshner, Robert P. and Matheson, Thomas and Sargent, Wallace L. W. and Simcoe, Robert A. and Smith, Graeme H. and Squires, Gordon},
	month = apr,
	year = {2003},
	note = {ADS Bibcode: 2003PASP..115..453L},
	pages = {453--473},
}

@article{blondin_using_2006,
	title = {Using {Line} {Profiles} to {Test} the {Fraternity} of {Type} {Ia} {Supernovae} at {High} and {Low} {Redshifts}*},
	volume = {131},
	issn = {1538-3881},
	url = {https://doi.org/10.1086/498724},
	doi = {10.1086/498724},
	language = {en},
	number = {3},
	urldate = {2026-02-14},
	journal = {The Astronomical Journal},
	author = {Blondin, Stéphane and Dessart, Luc and Leibundgut, Bruno and Branch, David and Höflich, Peter and Tonry, John L. and Matheson, Thomas and Foley, Ryan J. and Chornock, Ryan and Filippenko, Alexei V. and Sollerman, Jesper and Spyromilio, Jason and Kirshner, Robert P. and Wood-Vasey, W. Michael and Clocchiatti, Alejandro and Aguilera, Claudio and Barris, Brian and Becker, Andrew C. and Challis, Peter and Covarrubias, Ricardo and Davis, Tamara M. and Garnavich, Peter and Hicken, Malcolm and Jha, Saurabh and Krisciunas, Kevin and Li, Weidong and Miceli, Anthony and Miknaitis, Gajus and Pignata, Giuliano and Prieto, Jose Luis and Rest, Armin and Riess, Adam G. and Salvo, Maria Elena and Schmidt, Brian P. and Smith, R. Chris and Stubbs, Christopher W. and Suntzeff, Nicholas B.},
	month = mar,
	year = {2006},
	pages = {1648},
}

@article{filippenko_optical_1997,
	title = {{OPTICAL} {SPECTRA} {OF} {SUPERNOVAE}},
	volume = {35},
	issn = {0066-4146, 1545-4282},
	url = {https://www.annualreviews.org/content/journals/10.1146/annurev.astro.35.1.309},
	doi = {10.1146/annurev.astro.35.1.309},
	language = {en},
	number = {Volume 35, 1997},
	urldate = {2026-02-06},
	journal = {Annual Review of Astronomy and Astrophysics},
	publisher = {Annual Reviews},
	author = {Filippenko, Alexei V.},
	month = sep,
	year = {1997},
	pages = {309--355},
}

@article{howell_gemini_2005,
	title = {Gemini {Spectroscopy} of {Supernovae} from the {Supernova} {Legacy} {Survey}: {Improving} {High}-{Redshift} {Supernova} {Selection} and {Classification}},
	volume = {634},
	issn = {0004-637X},
	shorttitle = {Gemini {Spectroscopy} of {Supernovae} from the {Supernova} {Legacy} {Survey}},
	url = {https://iopscience.iop.org/article/10.1086/497119/meta},
	doi = {10.1086/497119},
	language = {en},
	number = {2},
	urldate = {2025-09-08},
	journal = {The Astrophysical Journal},
	publisher = {IOP Publishing},
	author = {Howell, D. A. and Sullivan, M. and Perrett, K. and Bronder, T. J. and Hook, I. M. and Astier, P. and Aubourg, E. and Balam, D. and Basa, S. and Carlberg, R. G. and Fabbro, S. and Fouchez, D. and Guy, J. and Lafoux, H. and Neill, J. D. and Pain, R. and Palanque-Delabrouille, N. and Pritchet, C. J. and Regnault, N. and Rich, J. and Taillet, R. and Knop, R. and McMahon, R. G. and Perlmutter, S. and Walton, N. A.},
	month = dec,
	year = {2005},
	pages = {1190},
}

@misc{sharma_ccsnscore_2025,
	title = {{CCSNscore}: {A} multi-input deep learning tool for classification of core-collapse supernovae using {SED}-{Machine} spectra},
	shorttitle = {{CCSNscore}},
	url = {http://arxiv.org/abs/2412.08601},
	doi = {10.48550/arXiv.2412.08601},
	urldate = {2025-09-08},
	publisher = {arXiv},
	author = {Sharma, Yashvi and Mahabal, Ashish A. and Sollerman, Jesper and Fremling, Christoffer and Kulkarni, S. R. and Rehemtulla, Nabeel and Miller, Adam A. and Aubert, Marie and Chen, Tracy X. and Coughlin, Michael W. and Graham, Matthew J. and Hale, David and Kasliwal, Mansi M. and Kim, Young-Lo and Neill, James D. and Purdum, Josiah N. and Rusholme, Ben and Singh, Avinash and Sravan, Niharika},
	month = mar,
	year = {2025},
	note = {arXiv:2412.08601 [astro-ph]},
}

@article{kim_how_2024,
	title = {How {Accurate} are {Transient} {Spectral} {Classification} {Tools}?— {A} {Study} {Using} 4646 {SEDMachine} {Spectra}},
	volume = {136},
	issn = {1538-3873},
	shorttitle = {How {Accurate} are {Transient} {Spectral} {Classification} {Tools}?},
	url = {https://dx.doi.org/10.1088/1538-3873/ad85cd},
	doi = {10.1088/1538-3873/ad85cd},
	language = {en},
	number = {11},
	urldate = {2025-09-08},
	journal = {Publications of the Astronomical Society of the Pacific},
	publisher = {The Astronomical Society of the Pacific},
	author = {Kim, Young-Lo and Hook, Isobel and Milligan, Andrew and Galbany, Lluís and Sollerman, Jesper and Burgaz, Umut and Dimitriadis, Georgios and Fremling, Christoffer and Johansson, Joel and Müller-Bravo, Tomás E. and Neill, James D. and Nordin, Jakob and Nugent, Peter and Purdum, Josiah and Qin, Yu-Jing and Rosnet, Philippe and Sharma, Yashvi},
	month = nov,
	year = {2024},
	pages = {114501},
}

@article{harutyunyan_esc_2008,
	title = {{ESC} supernova spectroscopy of non-{ESC} targets},
	volume = {488},
	copyright = {© ESO, 2008},
	issn = {0004-6361, 1432-0746},
	url = {https://www.aanda.org/articles/aa/abs/2008/34/aa8859-07/aa8859-07.html},
	doi = {10.1051/0004-6361:20078859},
	language = {en},
	number = {1},
	urldate = {2025-09-08},
	journal = {Astronomy \& Astrophysics},
	publisher = {EDP Sciences},
	author = {Harutyunyan, A. H. and Pfahler, P. and Pastorello, A. and Taubenberger, S. and Turatto, M. and Cappellaro, E. and Benetti, S. and Elias-Rosa, N. and Navasardyan, H. and Valenti, S. and Stanishev, V. and Patat, F. and Riello, M. and Pignata, G. and Hillebrandt, W.},
	month = sep,
	year = {2008},
	pages = {383--399},
}

@misc{vaswani_attention_2017,
	title = {Attention {Is} {All} {You} {Need}},
	url = {http://arxiv.org/abs/1706.03762},
	doi = {10.48550/arXiv.1706.03762},
	urldate = {2023-01-28},
	publisher = {arXiv},
	author = {Vaswani, Ashish and Shazeer, Noam and Parmar, Niki and Uszkoreit, Jakob and Jones, Llion and Gomez, Aidan N. and Kaiser, Lukasz and Polosukhin, Illia},
	month = dec,
	year = {2017},
	note = {arXiv:1706.03762 [cs]},
}

@article{modjaz_optical_2014,
	title = {{OPTICAL} {SPECTRA} {OF} 73 {STRIPPED}-{ENVELOPE} {CORE}-{COLLAPSE} {SUPERNOVAE}},
	volume = {147},
	issn = {1538-3881},
	url = {https://dx.doi.org/10.1088/0004-6256/147/5/99},
	doi = {10.1088/0004-6256/147/5/99},
	language = {en},
	number = {5},
	urldate = {2022-11-08},
	journal = {The Astronomical Journal},
	publisher = {The American Astronomical Society},
	author = {Modjaz, M. and Blondin, S. and Kirshner, R. P. and Matheson, T. and Berlind, P. and Bianco, F. B. and Calkins, M. L. and Challis, P. and Garnavich, P. and Hicken, M. and Jha, S. and Liu, Y. Q. and Marion, G. H.},
	month = mar,
	year = {2014},
	pages = {99},
}

@misc{liu_supernova_2015,
	title = {{SuperNova} {IDentification} spectral templates of 70 stripped-envelope core-collapse supernovae},
	url = {http://arxiv.org/abs/1405.1437},
	doi = {10.48550/arXiv.1405.1437},
	urldate = {2022-11-08},
	publisher = {arXiv},
	author = {Liu, Yuqian and Modjaz, Maryam},
	month = mar,
	year = {2015},
	note = {arXiv:1405.1437 [astro-ph]},
}

@article{matheson_optical_2008,
	title = {{OPTICAL} {SPECTROSCOPY} {OF} {TYPE} {Ia} {SUPERNOVAE}*},
	volume = {135},
	issn = {1538-3881},
	url = {https://dx.doi.org/10.1088/0004-6256/135/4/1598},
	doi = {10.1088/0004-6256/135/4/1598},
	language = {en},
	number = {4},
	urldate = {2022-11-08},
	journal = {The Astronomical Journal},
	publisher = {The American Astronomical Society},
	author = {Matheson, T. and Kirshner, R. P. and Challis, P. and Jha, S. and Garnavich, P. M. and Berlind, P. and Calkins, M. L. and Blondin, S. and Balog, Z. and Bragg, A. E. and Caldwell, N. and Concannon, K. Dendy and Falco, E. E. and Graves, G. J. M. and Huchra, J. P. and Kuraszkiewicz, J. and Mader, J. A. and Mahdavi, A. and Phelps, M. and Rines, K. and Song, I. and Wilkes, B. J.},
	month = mar,
	year = {2008},
	pages = {1598},
}

@article{yaron_wiserepinteractive_2012,
	title = {{WISeREP}—{An} {Interactive} {Supernova} {Data} {Repository}},
	volume = {124},
	issn = {1538-3873},
	url = {https://iopscience.iop.org/article/10.1086/666656/meta},
	doi = {10.1086/666656},
	language = {en},
	number = {917},
	urldate = {2022-11-08},
	journal = {Publications of the Astronomical Society of the Pacific},
	publisher = {IOP Publishing},
	author = {Yaron, Ofer and Gal-Yam, Avishay},
	month = jun,
	year = {2012},
	pages = {668},
}

@article{blondin_spectroscopic_2012,
	title = {{THE} {SPECTROSCOPIC} {DIVERSITY} {OF} {TYPE} {Ia} {SUPERNOVAE}*},
	volume = {143},
	issn = {1538-3881},
	url = {https://dx.doi.org/10.1088/0004-6256/143/5/126},
	doi = {10.1088/0004-6256/143/5/126},
	language = {en},
	number = {5},
	urldate = {2022-11-08},
	journal = {The Astronomical Journal},
	publisher = {The American Astronomical Society},
	author = {Blondin, S. and Matheson, T. and Kirshner, R. P. and Mandel, K. S. and Berlind, P. and Calkins, M. and Challis, P. and Garnavich, P. M. and Jha, S. W. and Modjaz, M. and Riess, A. G. and Schmidt, B. P.},
	month = apr,
	year = {2012},
	pages = {126},
}

@article{liu_analyzing_2016,
	title = {Analyzing the {Largest} {Spectroscopic} {Data} {Set} of {Stripped} {Supernovae} to {Improve} {Their} {Identifications} and {Constrain} {Their} {Progenitors}},
	volume = {827},
	issn = {0004-637X},
	url = {https://ui.adsabs.harvard.edu/abs/2016ApJ...827...90L},
	doi = {10.3847/0004-637X/827/2/90},
	urldate = {2022-11-08},
	journal = {The Astrophysical Journal},
	author = {Liu, Yu-Qian and Modjaz, Maryam and Bianco, Federica B. and Graur, Or},
	month = aug,
	year = {2016},
	note = {ADS Bibcode: 2016ApJ...827...90L},
	pages = {90},
}

@article{silverman_berkeley_2012,
	title = {Berkeley {Supernova} {Ia} {Program} – {I}. {Observations}, data reduction and spectroscopic sample of 582 low-redshift {Type} {Ia} supernovae},
	volume = {425},
	issn = {0035-8711},
	url = {https://doi.org/10.1111/j.1365-2966.2012.21270.x},
	doi = {10.1111/j.1365-2966.2012.21270.x},
	number = {3},
	urldate = {2022-11-08},
	journal = {Monthly Notices of the Royal Astronomical Society},
	author = {Silverman, Jeffrey M. and Foley, Ryan J. and Filippenko, Alexei V. and Ganeshalingam, Mohan and Barth, Aaron J. and Chornock, Ryan and Griffith, Christopher V. and Kong, Jason J. and Lee, Nicholas and Leonard, Douglas C. and Matheson, Thomas and Miller, Emily G. and Steele, Thea N. and Barris, Brian J. and Bloom, Joshua S. and Cobb, Bethany E. and Coil, Alison L. and Desroches, Louis-Benoit and Gates, Elinor L. and Ho, Luis C. and Jha, Saurabh W. and Kandrashoff, Michael T. and Li, Weidong and Mandel, Kaisey S. and Modjaz, Maryam and Moore, Matthew R. and Mostardi, Robin E. and Papenkova, Marina S. and Park, Sung and Perley, Daniel A. and Poznanski, Dovi and Reuter, Cassie A. and Scala, James and Serduke, Franklin J. D. and Shields, Joseph C. and Swift, Brandon J. and Tonry, John L. and Van Dyk, Schuyler D. and Wang, Xiaofeng and Wong, Diane S.},
	month = sep,
	year = {2012},
	pages = {1789--1818},
}

@techreport{lsst_science_collaboration_lsst_2009,
	title = {{LSST} {Science} {Book}, {Version} 2.0},
	url = {https://ui.adsabs.harvard.edu/abs/2009arXiv0912.0201L},
	urldate = {2022-11-08},
	author = {{LSST Science Collaboration} and Abell, Paul A. and Allison, Julius and Anderson, Scott F. and Andrew, John R. and Angel, J. Roger P. and Armus, Lee and Arnett, David and Asztalos, S. J. and Axelrod, Tim S. and Bailey, Stephen and Ballantyne, D. R. and Bankert, Justin R. and Barkhouse, Wayne A. and Barr, Jeffrey D. and Barrientos, L. Felipe and Barth, Aaron J. and Bartlett, James G. and Becker, Andrew C. and Becla, Jacek and Beers, Timothy C. and Bernstein, Joseph P. and Biswas, Rahul and Blanton, Michael R. and Bloom, Joshua S. and Bochanski, John J. and Boeshaar, Pat and Borne, Kirk D. and Bradac, Marusa and Brandt, W. N. and Bridge, Carrie R. and Brown, Michael E. and Brunner, Robert J. and Bullock, James S. and Burgasser, Adam J. and Burge, James H. and Burke, David L. and Cargile, Phillip A. and Chandrasekharan, Srinivasan and Chartas, George and Chesley, Steven R. and Chu, You-Hua and Cinabro, David and Claire, Mark W. and Claver, Charles F. and Clowe, Douglas and Connolly, A. J. and Cook, Kem H. and Cooke, Jeff and Cooray, Asantha and Covey, Kevin R. and Culliton, Christopher S. and de Jong, Roelof and de Vries, Willem H. and Debattista, Victor P. and Delgado, Francisco and Dell'Antonio, Ian P. and Dhital, Saurav and Di Stefano, Rosanne and Dickinson, Mark and Dilday, Benjamin and Djorgovski, S. G. and Dobler, Gregory and Donalek, Ciro and Dubois-Felsmann, Gregory and Durech, Josef and Eliasdottir, Ardis and Eracleous, Michael and Eyer, Laurent and Falco, Emilio E. and Fan, Xiaohui and Fassnacht, Christopher D. and Ferguson, Harry C. and Fernandez, Yanga R. and Fields, Brian D. and Finkbeiner, Douglas and Figueroa, Eduardo E. and Fox, Derek B. and Francke, Harold and Frank, James S. and Frieman, Josh and Fromenteau, Sebastien and Furqan, Muhammad and Galaz, Gaspar and Gal-Yam, A. and Garnavich, Peter and Gawiser, Eric and Geary, John and Gee, Perry and Gibson, Robert R. and Gilmore, Kirk and Grace, Emily A. and Green, Richard F. and Gressler, William J. and Grillmair, Carl J. and Habib, Salman and Haggerty, J. S. and Hamuy, Mario and Harris, Alan W. and Hawley, Suzanne L. and Heavens, Alan F. and Hebb, Leslie and Henry, Todd J. and Hileman, Edward and Hilton, Eric J. and Hoadley, Keri and Holberg, J. B. and Holman, Matt J. and Howell, Steve B. and Infante, Leopoldo and Ivezic, Zeljko and Jacoby, Suzanne H. and Jain, Bhuvnesh and {R} and {Jedicke} and Jee, M. James and Garrett Jernigan, J. and Jha, Saurabh W. and Johnston, Kathryn V. and Jones, R. Lynne and Juric, Mario and Kaasalainen, Mikko and {Styliani} and {Kafka} and Kahn, Steven M. and Kaib, Nathan A. and Kalirai, Jason and Kantor, Jeff and Kasliwal, Mansi M. and Keeton, Charles R. and Kessler, Richard and Knezevic, Zoran and Kowalski, Adam and Krabbendam, Victor L. and Krughoff, K. Simon and Kulkarni, Shrinivas and Kuhlman, Stephen and Lacy, Mark and Lepine, Sebastien and Liang, Ming and Lien, Amy and Lira, Paulina and Long, Knox S. and Lorenz, Suzanne and Lotz, Jennifer M. and Lupton, R. H. and Lutz, Julie and Macri, Lucas M. and Mahabal, Ashish A. and Mandelbaum, Rachel and Marshall, Phil and May, Morgan and McGehee, Peregrine M. and Meadows, Brian T. and Meert, Alan and Milani, Andrea and Miller, Christopher J. and Miller, Michelle and Mills, David and Minniti, Dante and Monet, David and Mukadam, Anjum S. and Nakar, Ehud and Neill, Douglas R. and Newman, Jeffrey A. and Nikolaev, Sergei and Nordby, Martin and O'Connor, Paul and Oguri, Masamune and Oliver, John and Olivier, Scot S. and Olsen, Julia K. and Olsen, Knut and Olszewski, Edward W. and Oluseyi, Hakeem and Padilla, Nelson D. and Parker, Alex and Pepper, Joshua and Peterson, John R. and Petry, Catherine and Pinto, Philip A. and Pizagno, James L. and Popescu, Bogdan and Prsa, Andrej and Radcka, Veljko and Raddick, M. Jordan and Rasmussen, Andrew and Rau, Arne and Rho, Jeonghee and Rhoads, James E. and Richards, Gordon T. and Ridgway, Stephen T. and Robertson, Brant E. and Roskar, Rok and Saha, Abhijit and Sarajedini, Ata and Scannapieco, Evan and Schalk, Terry and Schindler, Rafe and Schmidt, Samuel and Schmidt, Sarah and Schneider, Donald P. and Schumacher, German and Scranton, Ryan and Sebag, Jacques and Seppala, Lynn G. and Shemmer, Ohad and Simon, Joshua D. and Sivertz, M. and Smith, Howard A. and Allyn Smith, J. and Smith, Nathan and Spitz, Anna H. and Stanford, Adam and Stassun, Keivan G. and Strader, Jay and Strauss, Michael A. and Stubbs, Christopher W. and Sweeney, Donald W. and Szalay, Alex and Szkody, Paula and Takada, Masahiro and Thorman, Paul and Trilling, David E. and Trimble, Virginia and Tyson, Anthony and Van Berg, Richard and Vanden Berk, Daniel and VanderPlas, Jake and Verde, Licia and Vrsnak, Bojan and Walkowicz, Lucianne M. and Wandelt, Benjamin D. and Wang, Sheng and Wang, Yun and Warner, Michael and Wechsler, Risa H. and West, Andrew A. and Wiecha, Oliver and Williams, Benjamin F. and Willman, Beth and Wittman, David and Wolff, Sidney C. and Wood-Vasey, W. Michael and Wozniak, Przemek and Young, Patrick and Zentner, Andrew and Zhan, Hu},
	month = dec,
	year = {2009},
	note = {Publication Title: arXiv e-prints
ADS Bibcode: 2009arXiv0912.0201L
Type: article},
}

@article{athem_alsabti_handbook_2015,
	title = {Handbook of {Supernovae}},
	volume = {29},
	url = {https://ui.adsabs.harvard.edu/abs/2015IAUGA..2253696A},
	urldate = {2022-11-08},
	author = {Athem Alsabti, Abdul},
	month = aug,
	year = {2015},
	note = {Conference Name: IAU General Assembly
ADS Bibcode: 2015IAUGA..2253696A},
	pages = {2253696},
}

@article{couch_mechanisms_2017,
	title = {The mechanism(s) of core-collapse supernovae},
	volume = {375},
	url = {https://royalsocietypublishing.org/doi/10.1098/rsta.2016.0271},
	doi = {10.1098/rsta.2016.0271},
	number = {2105},
	urldate = {2022-11-08},
	journal = {Philosophical Transactions of the Royal Society A: Mathematical, Physical and Engineering Sciences},
	publisher = {Royal Society},
	author = {Couch, Sean M.},
	month = oct,
	year = {2017},
	pages = {20160271},
}

@article{riess_observational_1998,
	title = {Observational {Evidence} from {Supernovae} for an {Accelerating} {Universe} and a {Cosmological} {Constant}},
	volume = {116},
	issn = {1538-3881},
	url = {https://dx.doi.org/10.1086/300499},
	doi = {10.1086/300499},
	language = {en},
	number = {3},
	urldate = {2022-11-08},
	journal = {The Astronomical Journal},
	author = {Riess, Adam G. and Filippenko, Alexei V. and Challis, Peter and Clocchiatti, Alejandro and Diercks, Alan and Garnavich, Peter M. and Gilliland, Ron L. and Hogan, Craig J. and Jha, Saurabh and Kirshner, Robert P. and Leibundgut, B. and Phillips, M. M. and Reiss, David and Schmidt, Brian P. and Schommer, Robert A. and Smith, R. Chris and Spyromilio, J. and Stubbs, Christopher and Suntzeff, Nicholas B. and Tonry, John},
	month = sep,
	year = {1998},
	pages = {1009},
}

@article{perlmutter_measurements_1999,
	title = {Measurements of Ω and Λ from 42 {High}-{Redshift} {Supernovae}},
	volume = {517},
	issn = {0004-637X},
	url = {https://dx.doi.org/10.1086/307221},
	doi = {10.1086/307221},
	language = {en},
	number = {2},
	urldate = {2022-11-08},
	journal = {The Astrophysical Journal},
	author = {Perlmutter, S. and Aldering, G. and Goldhaber, G. and Knop, R. A. and Nugent, P. and Castro, P. G. and Deustua, S. and Fabbro, S. and Goobar, A. and Groom, D. E. and Hook, I. M. and Kim, A. G. and Kim, M. Y. and Lee, J. C. and Nunes, N. J. and Pain, R. and Pennypacker, C. R. and Quimby, R. and Lidman, C. and Ellis, R. S. and Irwin, M. and McMahon, R. G. and Ruiz-Lapuente, P. and Walton, N. and Schaefer, B. and Boyle, B. J. and Filippenko, A. V. and Matheson, T. and Fruchter, A. S. and Panagia, N. and Newberg, H. J. M. and Couch, W. J. and Project, The Supernova Cosmology},
	month = jun,
	year = {1999},
	pages = {565},
}

@article{minkowski_spectra_1941,
	title = {Spectra of {Supernovae}},
	volume = {53},
	issn = {0004-6280},
	url = {https://ui.adsabs.harvard.edu/abs/1941PASP...53..224M},
	doi = {10.1086/125315},
	urldate = {2022-11-08},
	journal = {Publications of the Astronomical Society of the Pacific},
	author = {Minkowski, R.},
	month = aug,
	year = {1941},
	note = {ADS Bibcode: 1941PASP...53..224M},
	pages = {224},
}

@article{modjaz_spectral_2016,
	title = {{THE} {SPECTRAL} {SN}-{GRB} {CONNECTION}: {SYSTEMATIC} {SPECTRAL} {COMPARISONS} {BETWEEN} {TYPE} {Ic} {SUPERNOVAE} {AND} {BROAD}-{LINED} {TYPE} {Ic} {SUPERNOVAE} {WITH} {AND} {WITHOUT} {GAMMA}-{RAY} {BURSTS}},
	volume = {832},
	issn = {0004-637X},
	shorttitle = {{THE} {SPECTRAL} {SN}-{GRB} {CONNECTION}},
	url = {https://dx.doi.org/10.3847/0004-637X/832/2/108},
	doi = {10.3847/0004-637X/832/2/108},
	language = {en},
	number = {2},
	urldate = {2022-11-07},
	journal = {The Astrophysical Journal},
	publisher = {The American Astronomical Society},
	author = {Modjaz, Maryam and Liu, Yuqian Q. and Bianco, Federica B. and Graur, Or},
	month = nov,
	year = {2016},
	pages = {108},
}

@article{blagorodnova_sed_2018,
	title = {The {SED} {Machine}: {A} {Robotic} {Spectrograph} for {Fast} {Transient} {Classification}},
	volume = {130},
	issn = {0004-6280},
	shorttitle = {The {SED} {Machine}},
	url = {https://ui.adsabs.harvard.edu/abs/2018PASP..130c5003B},
	doi = {10.1088/1538-3873/aaa53f},
	urldate = {2022-11-07},
	journal = {Publications of the Astronomical Society of the Pacific},
	author = {Blagorodnova, Nadejda and Neill, James D. and Walters, Richard and Kulkarni, Shrinivas R. and Fremling, Christoffer and Ben-Ami, Sagi and Dekany, Richard G. and Fucik, Jason R. and Konidaris, Nick and Nash, Reston and Ngeow, Chow-Choong and Ofek, Eran O. and O' Sullivan, Donal and Quimby, Robert and Ritter, Andreas and Vyhmeister, Karl E.},
	month = mar,
	year = {2018},
	note = {ADS Bibcode: 2018PASP..130c5003B},
	pages = {035003},
}

@article{fremling_sniascore_2021,
	title = {{SNIascore}: {Deep}-learning {Classification} of {Low}-resolution {Supernova} {Spectra}},
	volume = {917},
	issn = {0004-637X},
	shorttitle = {{SNIascore}},
	url = {https://ui.adsabs.harvard.edu/abs/2021ApJ...917L...2F},
	doi = {10.3847/2041-8213/ac116f},
	urldate = {2022-11-07},
	journal = {The Astrophysical Journal},
	author = {Fremling, Christoffer and Hall, Xander J. and Coughlin, Michael W. and Dahiwale, Aishwarya S. and Duev, Dmitry A. and Graham, Matthew J. and Kasliwal, Mansi M. and Kool, Erik C. and Mahabal, Ashish A. and Miller, Adam A. and Neill, James D. and Perley, Daniel A. and Rigault, Mickael and Rosnet, Philippe and Rusholme, Ben and Sharma, Yashvi and Shin, Kyung Min and Shupe, David L. and Sollerman, Jesper and Walters, Richard S. and Kulkarni, S. R.},
	month = aug,
	year = {2021},
	note = {ADS Bibcode: 2021ApJ...917L...2F},
	pages = {L2},
}

@article{williamson_optimal_2019,
	title = {Optimal {Classification} and {Outlier} {Detection} for {Stripped}-envelope {Core}-collapse {Supernovae}},
	volume = {880},
	issn = {0004-637X},
	url = {https://ui.adsabs.harvard.edu/abs/2019ApJ...880L..22W},
	doi = {10.3847/2041-8213/ab2edb},
	urldate = {2022-11-07},
	journal = {The Astrophysical Journal},
	author = {Williamson, Marc and Modjaz, Maryam and Bianco, Federica B.},
	month = aug,
	year = {2019},
	note = {ADS Bibcode: 2019ApJ...880L..22W},
	pages = {L22},
}

@article{muthukrishna_dash_2019,
	title = {{DASH}: {Deep} {Learning} for the {Automated} {Spectral} {Classification} of {Supernovae} and {Their} {Hosts}},
	volume = {885},
	issn = {0004-637X},
	shorttitle = {{DASH}},
	url = {https://ui.adsabs.harvard.edu/abs/2019ApJ...885...85M},
	doi = {10.3847/1538-4357/ab48f4},
	urldate = {2022-11-07},
	journal = {The Astrophysical Journal},
	author = {Muthukrishna, Daniel and Parkinson, David and Tucker, Brad E.},
	month = nov,
	year = {2019},
	note = {ADS Bibcode: 2019ApJ...885...85M},
	pages = {85},
}

@article{tonry_survey_1979,
	title = {A survey of galaxy redshifts. {I}. {Data} reduction techniques.},
	volume = {84},
	issn = {0004-6256},
	url = {https://ui.adsabs.harvard.edu/abs/1979AJ.....84.1511T},
	doi = {10.1086/112569},
	urldate = {2022-11-07},
	journal = {The Astronomical Journal},
	author = {Tonry, J. and Davis, M.},
	month = oct,
	year = {1979},
	note = {ADS Bibcode: 1979AJ.....84.1511T},
	pages = {1511--1525},
}

@article{blondin_determining_2007,
	title = {Determining the {Type}, {Redshift}, and {Age} of a {Supernova} {Spectrum}},
	volume = {666},
	issn = {0004-637X},
	url = {https://ui.adsabs.harvard.edu/abs/2007ApJ...666.1024B},
	doi = {10.1086/520494},
	urldate = {2022-11-07},
	journal = {The Astrophysical Journal},
	author = {Blondin, Stéphane and Tonry, John L.},
	month = sep,
	year = {2007},
	note = {ADS Bibcode: 2007ApJ...666.1024B},
	pages = {1024--1047},
}
\bibliographystyle{aasjournal}



\end{document}